\documentclass[prd,superscriptaddress,showpacs,nofootinbib,amsmath,amssymb]{revtex4}

\usepackage{bm}
\usepackage{amsfonts}
\usepackage{latexsym}
\usepackage[latin1]{inputenc}
\usepackage{graphicx}
\usepackage{amsmath}
\usepackage{rotating}
\usepackage{epsfig}
\usepackage{subfig}
\makeatletter 
\renewcommand{\p@subfigure}{} 
\makeatother 

\def \nn  {\nonumber}

\def\jnl@style{\it}
\def\aaref@jnl#1{{\jnl@style#1}}
\def\aaref@jnl#1{{\jnl@style#1}}

\def\aj{\aaref@jnl{AJ}}                   % Astronomical Journal
\def\apj{\aaref@jnl{ApJ}}                 % Astrophysical Journal
\def\apjl{\aaref@jnl{ApJ}}                % Astrophysical Journal, Letters
\def\apjs{\aaref@jnl{ApJS}}               % Astrophysical Journal, Supplement
\def\apss{\aaref@jnl{Ap\&SS}}             % Astrophysics and Space Science
\def\aap{\aaref@jnl{A\&A}}                % Astronomy and Astrophysics
\def\aapr{\aaref@jnl{A\&A~Rev.}}          % Astronomy and Astrophysics Reviews
\def\aaps{\aaref@jnl{A\&AS}}              % Astronomy and Astrophysics, Supplement
\def\mnras{\aaref@jnl{MNRAS}}             % Monthly Notices of the RAS
\def\prd{\aaref@jnl{Phys.~Rev.~D}}        % Physical Review D
\def\prl{\aaref@jnl{Phys.~Rev.~Lett.}}    % Physical Review Letters
\def\qjras{\aaref@jnl{QJRAS}}             % Quarterly Journal of the RAS
\def\skytel{\aaref@jnl{S\&T}}             % Sky and Telescope
\def\ssr{\aaref@jnl{Space~Sci.~Rev.}}     % Space Science Reviews
\def\zap{\aaref@jnl{ZAp}}                 % Zeitschrift fuer Astrophysik
\def\nat{\aaref@jnl{Nature}}              % Nature
\def\aplett{\aaref@jnl{Astrophys.~Lett.}} % Astrophysics Letters
\def\apspr{\aaref@jnl{Astrophys.~Space~Phys.~Res.}} % Astrophysics Space Physics Research
\def\physrep{\aaref@jnl{Phys.~Rep.}}      % Physics Reports
\def\physscr{\aaref@jnl{Phys.~Scr}}       % Physica Scripta

\begin{document}

%%%%%%%%%%%%%%%%%%%%%%%%%%%%%%%  TITLE  %%%%%%%%%%%%%%%%%%%%%%%%%%%%%%%%%%%%%%%%
\title[Fast Rotating Stars]{Oscillations of rapidly rotating relativistic stars}

%%%%%%%%%%%%%%%%%%%%%%%%%%%%%%%%  AUTHORS  %%%%%%%%%%%%%%%%%%%%%%%%%%%%%%%%%%%%%

\author{Erich Gaertig} \email{gaertig@tat.physik.uni-tuebingen.de}
\affiliation{Theoretical Astrophysics, University of T\"ubingen, Auf der Morgenstelle 10, T\"ubingen 72076, Germany}
\affiliation{Max-Planck-Institut f\"ur Gravitationsphysik, Albert-Einstein-Institut, Hannover 30167, Germany}

\author{Kostas D. Kokkotas} % \email{kokkotas@auth.gr}
\affiliation{Theoretical Astrophysics, University of T\"ubingen, Auf der Morgenstelle 10, T\"ubingen 72076, Germany}
\affiliation{Department of Physics,Aristotle University of Thessaloniki, Thessaloniki 54124, Greece}
%\affiliation{Department of Physics, Aristotle University of Thessaloniki, 54124, Greece}
%%%%%%%%%%%%%%%%%%%%%%%%%%%%%%%%%%%%  DATE  %%%%%%%%%%%%%%%%%%%%%%%%%%%%%%%%%%%%
\date{\today}

%%%%%%%%%%%%%%%%%%%%%%%%%%%%%  ABSTRACT  %%%%%%%%%%%%%%%%%%%%%%%%%%%%%%%%%%%%%%%

\begin{abstract}
Non-axisymmetric oscillations of rapidly rotating relativistic stars are studied using the Cowling approximation. The oscillation spectra have been estimated by  Fourier transforming the evolution equations describing the perturbations.
This is the first study of its kind and provides information on the effect of fast rotation on the oscillation spectra while it offers the possibility in studying the complete problem by including spacetime perturbations. 
Our study includes both axisymmetric and non-axisymmetric perturbations and provides limits for the onset of the secular bar mode rotational instability.
We also present approximate formulae for the dependence of the oscillation spectrum from rotation. The results suggest that it is possible to  extract  the relativistic star's parameters from the observed gravitational wave spectrum.

\end{abstract}

%%%%%%%%%%%%%%%%%%%%%%%%%%%%%  PACS  %%%%%%%%%%%%%%%%%%%%%%%%%%%%%%%%%%%%%%%%%%%
\pacs{04.30.Db, 04.40.Dg, 95.30.Sf, 97.10.Sj}

%%%%%%%%%%%%%%%%%%%%%%%%%%%%%  MAKETITLE  %%%%%%%%%%%%%%%%%%%%%%%%%%%%%%%%%%%%%%
\maketitle

%%%%%%%%%%%%%%%%%%%%%%%%%  SEC. I: INTRODUCTION  %%%%%%%%%%%%%%%%%%%%%%%%%%%%%%%
\section{Introduction}
%%%%%%%%%%%%%%%%%%%%%%%%%
Relativistic stars during  their evolution may undergo oscillations which can become unstable under certain conditions. Newly born neutron stars are expected to oscillate wildly during their creation shortly after the supernovae collapse \cite{Ott:2006eh}. They are expected also to oscillate if they are members of binary systems  and there is tidal interaction \cite{Kokkotas:1995xe,Flanagan:2006sb} or mass and angular momentum transfer from a companion star and also when they undergo phase transitions \cite{2002PhRvD..66f4027M,Ferrari:2007ur} which might be responsible for the observed glitches in pulsars. Rotation  strongly affects these oscillations and perturbed stars can become unstable if they rotate faster than some critical velocities. During these oscillatory phases of their lives compact stars emit copious amounts of gravitational waves which together with viscosity tend to suppress the oscillations. The oscillations are divided into different families according to the restoring force \cite{1988ApJ...325..725M,1999LRR.....2....2K}. If pressure is the main restoring force then these modes are called (pressure) \emph{p-modes}, buoyancy is the restoring force of another class of oscillation mode, the \emph{g-modes} while Coriolis force is the restoring force for the \emph{inertial modes}. Spacetime induces another family of oscillations which couple only weakly to the fluid, these are the so-called \emph{w-modes} \cite{Kokkotas:1992ks}. There are more families of modes if one assumes the presence of crust \cite{1983MNRAS.203..457S,1985ApJ...297L..37M,1988ApJ...325..725M,2008MNRAS.384.1711V} or magnetic fields \cite{1986ApJ...305..767C}. For a complete description of the relativistic star perturbation theory one may refer to \cite{1999LRR.....2....2K,2001IJMPD..10..381A,lrr-2003-3}.

The study of stellar perturbations in the framework of general relativity (GR) dates back to mid 1960s with the seminal works by Thorne and his collaborators \cite{Thorne:1967th,Price:1969pt,Thorne:1969to,Thorne:1969th}. Since then the study of stellar oscillations and possible instabilities was a field of intensive work in  relativistic astrophysics \cite{Detweiler:1985dl, 1985ApJ...297L..37M}. Moreover, during the last two decades, these studies become even more important due to the relations of the oscillations and instabilities to the emission of gravitational waves and the possibility of getting information about the stellar parameters (mass, radius, equation of state) by the proper analysis of the oscillation spectrum \cite{Andersson:1996ak,Andersson:1998ak,Benhar:1998au,2001MNRAS.320...307K,benhar-2004-70,2004PhRvD..69h4008S,2004PhRvD..70h4026S}. Still, all these studies were mainly dealing with non-rotating stars, because the combination of rotation and general relativity made both the analytic and numerical studies extremely involved. This led to certain approximations in studying rotating stars in GR. The most obvious of them include the so called ``slow-rotation" and the ``Cowling" approximation. Actually, both approximation were known and have been used extensively in Newtonian theory of stellar oscillations \cite{1989nos..book.....U}. In the \emph{slow rotation approximation} one expands the perturbation equations in terms of a small parameter $\varepsilon=\Omega/ \Omega_K$, where $\Omega$ is the angular velocity of the star and $\Omega_K$ stands for the ``Kepler angular velocity" which is the maximum velocity that can be attained before the star splits apart due to rotation. In the \emph{Cowling approximation} one typically neglects the perturbations of the Newtonian potential or the spacetime in the case of GR. This is a quite good approximation, both qualitatively and quantitatively, for the higher order \emph{p-modes}, for the \emph{g-modes} and the inertial modes while  it is only qualitatively  good  for the \emph{f-modes}, see for example \cite{1968AnAp...31..549R}. 
Although, most of our understanding on the oscillations of relativistic stars is due to perturbative studies, recently,  it has become possible to study stellar oscillations using evolutions of the non-linear equations of motion for the fluid \cite{Font:1999wh,2001PhRvL..86.1148S,Font:2001eu,Stergioulas:2003ep,Dimmelmeier:2005zk,2006PhRvD..74l4024K,2008arXiv0804.1151K,2008PhRvD..77d4042B,2008arXiv0803.3804B}. Finally, differential rotation is another key issue that is believed to play an important role in the dynamics of nascent neutron stars. Actually, it is associated with dynamical instabilities both for fast and slowly rotating neutron stars \cite{2001ApJ...550L.193C,2002MNRAS.334L..27S} and affects the onset of secular instabilities; still it has not yet been studied extensively  \cite{Dimmelmeier:2002bm,2007PhRvD..75f4019S,2008PhRvD..77b4029P} and remains an open issue.

Since rotational instabilities are typically connected with fast rotating stars, it is of the great importance the study of oscillations of this type of objects. As a first step one can drop the ``slow rotation" approximation but still maintain the Cowling approximation i.e. to freeze the spacetime perturbations or in the best case to freeze the radiative part of these perturbations. This was the approach used up to now for most of the studies of the oscillations of fast rotating relativistic stars either in perturbative approaches \cite{2002ApJ...568L..41Y,2007PhRvD..75d3007B} or in non-linear (but axisymmetric) approaches \cite{2006MNRAS.368.1609D,2008arXiv0804.1151K}.

In this article, we present the first results of a new approach based on 2D time evolution of the perturbation equations which seems to be promising for the study of the oscillations and instabilities of fast rotating neutron stars.  This the first study of its kind, since earlier 2D perturbative studies have been done either in Newtonian theory \cite{2002MNRAS.334..933J} (see also \cite{Passamonti:2008nx} for a very recent work) or in reduction to eigenvalue problem \cite{2002ApJ...568L..41Y,2007PhRvD..75d3007B,2007PhRvD..76j4033F}. The advantage of this method is that it can be easily extended to include differential rotation or the perturbations of the spacetime. On the other hand it provides a robust tool in studying the onset of rotational instabilities in fast rotating neutron stars, while as it has been demonstrated here, one can get easily results for realist equations of state which is vital for developing gravitational wave asteroseismology. 
Finally, via this approach one may answer the question of the existence or absence of a continuous spectrum for the inertial modes as it has been suggested by the 1D studies in the slow rotation approximation \cite{1998MNRAS.293...49K,Beyer:1999te,2002MNRAS.330.1027R,2003MNRAS.339.1170R,2008PhRvD..77b4029P}.

In the next section we describe in detail the derivation of the perturbation equations and the conventions that we have adopted. In section 3, we present the numerical tools that have been developed in order to study the problem while section 4  describes the results for axisymmetric and non-axisymmetric perturbations. In the last section we discuss the results, their application to astrophysics and the possible extensions of this work.

%

%%%%%%%%%%%%%%%%%%%%%%%%%  SEC. II: EQUATIONS  %%%%%%%%%%%%%%%%%%%%%%%%%%%%%%%

\section{The perturbation equations}
\label{sec:pertEq}
The study of oscillations of rotating neutron stars involves the solution of the full nonlinear set of Einstein's equations of General Relativity together with the equations of motion for the fluid (we set $G=c=1$ here)
\begin{equation}
\label{eq:einstein1}
G_{\mu\nu}  = 8\pi T_{\mu\nu} 
\end{equation}
\begin{equation}
\nabla_{\mu}T^{\mu\nu} = 0
 \label{eq:einstein2}, 
\end{equation}
where $G_{\mu\nu}$ is the Einstein-tensor, describing the geometry of space-time, and $T_{\mu\nu}$ is the energy-momentum--tensor that defines the functional form of energy momentum and stress of the fluid. Since it is very complicated, but not impossible \cite{2006MNRAS.368.1609D,2008arXiv0804.1151K},
 to solve this system as such, we will introduce some approximations in studying the problem. 

First of all, we linearize  equations (\ref{eq:einstein1}) and (\ref{eq:einstein2}), which means that we constrain our study to small perturbations around the equilibrium. Secondly, we will work in the so-called Cowling-approximation, which means that we will neglect all metric perturbations. This simplifies significantly  the equations one has to solve since  the space-time is considered as ``frozen''; and we only have to solve the linearized version  of \eqref{eq:einstein2}. 
Under this assumption equation \eqref{eq:einstein2} will be written as
%In the following, we will write the quantities that perturbation quantities $A$ as
%\begin{equation}
%\label{eq:conventionForPerturbations}
%\delta A = A_{0} + \delta A\,,
%\end{equation}
%where $A_{0}$ is the unperturbed background quantity and $\delta A$ its linear perturbation. With this convention we have to study the solution of
\begin{equation}
\label{eq:perturbedEinstein}
\nabla_{\mu}(\delta T^{\mu\nu}) = g^{\mu\kappa}(\partial_{\mu}\delta T_{\kappa\nu}-{\Gamma^{\lambda}}_{\kappa\mu}\delta T_{\lambda\nu}
-{\Gamma^{\lambda}}_{\mu\nu}\delta T_{\kappa\lambda}) = 0
\end{equation} 
where $g^{\mu\nu}$ is the metric describing neutron star's spacetime, ${\Gamma^{\lambda}}_{\kappa\mu}$ are the Christoffel symbols and $\delta T_{\mu\nu}$  is the Eulerian perturbation of the energy-momentum--tensor. We assume that the  matter has no viscosity or shear stresses, i.e. that it can be described by a perfect fluid.  Thus $\delta T_{\mu\nu}$ has the form
\begin{equation}
\label{eq:deltaTmunu}
\delta T_{\mu\nu}=(\epsilon+p)(u_{\mu}\delta u_{\nu}+u_{\nu}\delta u_{\mu})+(\delta p+\delta\epsilon)u_{\mu}u_{\nu}+\delta p g_{\mu\nu}\,.
\end{equation}
where $\epsilon$ is the energy-density, $p$ is the pressure, $u_{\mu}$ is the 4-velocity and $\delta u_{\mu}$ are its perturbation. Energy density and pressure are not independent quantities but are connected via an equation of state (EOS) which we assume to be polytropic, i.e.
\begin{eqnarray}
\label{eq:polytropicEOS}
p &=& K\tilde{\rho}^{1 + 1/N}\quad \mbox{where} \quad
\epsilon = \tilde{\rho} +Np\,.
\end{eqnarray}
Here $\tilde{\rho}$ is the rest-mass density, $K$ the polytropic constant, $N$ the polytropic exponent and $\Gamma = 1 + 1/N$ the polytropic index. For barotropic oscillations, both the unperturbed background and the perturbations are described by the same equation of state. In this case the pressure variation can be replaced by the corresponding energy-density variation via $\delta p = c_{s}^{2}\delta\epsilon$, where $c_s$ is  the speed of sound 
\begin{equation}
\label{eq:soundSpeed}
c_{s}^{2} = \frac{\partial p}{\partial\epsilon}\, .
\end{equation}
For polytropic EOS of the form \eqref{eq:polytropicEOS} it is given by
\begin{equation}
\label{eq:polytropicSoundSpeed}
c_{s}^{2} = \frac{\Gamma p}{\epsilon + p}\,.
\end{equation}
Our background model is a compact relativistic star that rotates uniformly up to its Kepler-limit, i.e. the point were it is torn apart by centrifugal forces. In this work we will adopt the metric in a comoving frame of reference as it is described in \cite{Ansorg:2003mz}. In Lewis-Papapetrou coordinates $(\rho, \zeta, \varphi,t)$ the metric reads
\begin{equation}
\label{eq:lineElement}
ds^{2} = e^{-2U}\left[e^{2k}\left(d\rho^{2}+d\zeta^{2}\right)+W^{2}d\varphi^{2}\right]-e^{2U}(dt+a d\varphi)^{2}\,,
\end{equation}
%\begin{equation}
%\label{eq:metric}
%g_{\mu\nu}=\left(\begin{array}{cccc}
% e^{-2U+2k} & 0 & 0 &0\\
%0 & e^{-2U+2k} & 0 & 0\\
%0 & 0 & W^{2}e^{-2U}-a^{2}e^{2U} & -a\,e^{2U}\\
%0 & 0 & -a\,e^{2U} & -e^{2U}\\
%\end{array}\right)\,,
%\end{equation}
where all functions depend on $\rho$ and $\zeta$. In general, the $(\varphi,t)$-metric component is proportional to the function $a$ and vanishes in the absence of rotation. Other properties of the metric functions which will become important later are
\begin{eqnarray}
\label{eq:limes}
\lim_{\rho\to 0}|a\rho^{-2}| &<& \infty\quad \mbox{and} \quad
\lim_{\rho\to 0}|W\rho^{-1}| < \infty\, .
\end{eqnarray}
The utilization of a comoving frame of reference as implemented in  \cite{Ansorg:2003mz} has several important advantages. First of all is the use of surface-fitted coordinates. Especially when the star is rapidly rotating, deviations from spherical symmetry become more and more evident. If one uses a fixed spherical symmetric grid to describe the neutron star, even if the boundary of the nonrotating configuration can be aligned to a single grid line, this will fail once the star rotates and gets deformed, so that the surface will lie somewhere between different grid lines. This may cause problems when implementing boundary conditions at the surface. With surface-fitting coordinates, even the surface of the most rapidly rotating configuration can be described by a single parameter.
Secondly, comoving coordinates reduce the complexity and length of the equations to solve considerably. Let us briefly review how one would proceed in a stationary frame: The first observation is, that there are still dependent quantities in the equations \eqref{eq:perturbedEinstein} and \eqref{eq:deltaTmunu}. One would write down the definition of the constant angular velocity
\begin{equation*}
\frac{u^{\varphi}}{u^{t}} = \Omega \hspace{0.5cm}\mbox{\rm{and use the relationship}}\hspace{0.5cm}g_{\mu\nu}u^{\mu}u^{\nu} = -1
\end{equation*}
to get
%\begin{equation*}
$
u^{t} = (-g_{tt} - 2\Omega g_{\varphi t} -\Omega^{2}g_{\varphi\varphi})^{-1/2}
$.
%\,.
%\end{equation*}
This is how the angular velocity enters the equations explicitly. Furthermore since also
\begin{equation*}
g_{\mu\nu}(u^{\mu} + \delta u^{\mu})(u^{\nu} + \delta u^{\nu}) = -1\,, \hspace{0.5cm}\mbox{\rm{we have}}\hspace{0.5cm}\delta u_{t} = -\Omega\delta u_{\varphi}\,.
\end{equation*}
In this way we can write both $u^{t}$ and $\delta u^{t}$ as functions of the angular velocity. On the other hand, in a comoving frame $u^{\varphi} = 0$ while the equation for $\delta u_{t}$ is trivial; i.e. $\delta u_{t} = 0$. This reduces the equations governing linear perturbations considerably as we will see now.

For non-rotating or slowly rotating stars where the background configuration was considered as spherical the perturbation equations were decomposed into spherical harmonics and the problem was typically reduced in solving the equations describing only the radial components of the perturbations. But here we deal with fast rotating neutron stars which are deformed due to rotation, this means that it is no longer possible to decompose the angular part of our perturbation quantitites in spherical harmonics as it was usually done for the non-rotating case (see \cite{Thorne:1967th}) or in the slow-rotation-approximation (for example \cite{Kojima:1992kj} and \cite{2005IJMPD..14..543S}). Instead we can only separate the azimuthal dependence and the perturbation functions that we use will be written as
\begin{eqnarray}
\label{eq:perturbedAnsatz}
(\epsilon + p)W e^{U}\,\delta u_{\rho}&=&f_{1}(\rho,\zeta,t)e^{im\varphi}\nn\\
(\epsilon + p)W e^{U}\,\delta u_{\zeta}&=&f_{2}(\rho,\zeta,t)e^{im\varphi}\\
(\epsilon + p)\,\delta u_{\varphi}&=&f_{3}(\rho,\zeta,t)e^{im\varphi}\nn\\
c_{s}^{2}e^{U}\,\delta\epsilon&=&H(\rho,\zeta,t)e^{im\varphi}\nn \, .
\end{eqnarray}
The functions $f_{i}\,,i=1\ldots 3$ are describing the time-evolution of the perturbed velocity components and $H$ describes the corresponding change in energy-density. All these functions are in general complex-valued (of course not for the initial starting condition at $t=t_{0}$) and multiplied by a complex phase that prescribes the dependence on the azimuthal angle $\varphi$. This means, that we will get a set of complex-valued PDEs and since this is a linear system, the final solution is obtained by taking the real part of these quantitities.\\
%The resulting equations can be simplified quite a lot by the substitution
%\begin{eqnarray}
%\label{eq:newVariables}
%\tilde{H} &=& c_{s}^{2}e^{U}H\\
%\tilde{f}_{1} &=& (\epsilon + p)W e^{U}f_{1}\nn\\
%\tilde{f}_{2} &=& (\epsilon + p)W e^{U}f_{2}\nn\\
%\tilde{f}_{3} &=& (\epsilon + p)f_{3}\nn
%\end{eqnarray}
The substitution of the  perturbation functions \eqref{eq:perturbedAnsatz} into equations \eqref{eq:perturbedEinstein}, leads to the following  system of evolution equations:
\begin{eqnarray}
\label{eq:explicit_v2}
\frac{\partial{f}_{1}}{\partial t}& = &-We^{U}\frac{\partial{H}}{\partial \rho}
-\frac{e^{5U}}{W}\frac{\partial a}{\partial \rho}{f}_{3}
-\frac{W}{c_{s}^{2}}\frac{\partial U}{\partial \rho}e^{U}{H} \nn \\
\frac{\partial{f}_{2}}{\partial t}& = &-We^{U}\frac{\partial{H}}{\partial\zeta}
-\frac{e^{5U}}{W}\frac{\partial a}{\partial \zeta}{f}_{3}
-\frac{W}{c_{s}^{2}}\frac{\partial U}{\partial \zeta}e^{U}{H}\nn\\
\frac{\partial{f}_{3}}{\partial t}& = &\frac{im}{F}\left(ac_{s}^{2}e^{4U}{f}_{3}+W^{2}{H}\right)
+\frac{Wac_{s}^{2}e^{3U-2k}}{F}\left(\frac{\partial{f}_{1}}{\partial \rho}+\frac{\partial{f}_{2}}{\partial \zeta}\right)\\
& &-\frac{e^{3U-2k}}{F}W\frac{\partial a}{\partial \rho}{f}_{1}
-\frac{e^{3U-2k}}{F}W\frac{\partial a}{\partial \zeta}{f}_{2}\nn\\
\frac{\partial{H}}{\partial t}& = &\frac{im}{F}\left(ac_{s}^{2}e^{4U}{H}+c_{s}^{2}e^{4U}{f}_{3}\right)
+\frac{Wc_{s}^{2}e^{3U-2k}}{F}\left(\frac{\partial{f}_{1}}{\partial \rho}+\frac{\partial{f}_{2}}{\partial \zeta}\right)\nn\\
& & -\frac{c_{s}^{2}e^{7U-2k}}{F}\frac{a}{W}\frac{\partial a}{\partial \rho}{f}_{1}-\frac{c_{s}^{2}e^{7U-2k}}{F}\frac{a}{W}\frac{\partial a}{\partial \zeta}{f}_{2}\nn
\end{eqnarray}
where
\begin{equation}
\label{eq:Fdefinition}
F:=a^{2}c_{s}^{2}e^{4U}-W^{2}\,.
\end{equation}
As we discussed earlier, there is no explicit dependence on the angular velocity $\Omega$ in this system of equations. %This means that the structure of the system remains unchanged by the rotation rate.

%One can simplify this system even more by combining $f_{3}$ and $H$ into a new function. Then the last two equations of \eqref{eq:explicit_v2} can be written as
%\begin{eqnarray}
%\label{eq:linearComb}
%W\frac{\partial{f}_{3}}{\partial t}-Wa\frac{\partial{H}}{\partial t}&=&-imW{H}+e^{3U-2k}\left(\frac{\partial a}{\partial\rho}{f}_{1}+
%\frac{\partial a}{\partial \zeta}{f}_{2}\right)\\
%a\frac{\partial{f}_{3}}{\partial t}-\frac{W^{2}}{c_{s}^{2}e^{4U}}\frac{\partial{H}}{\partial t}&=&im{f}_{3}+We^{-U-2k}\left(\frac{\partial{f}_{1}}{\partial \rho}+\frac{\partial{f}_{2}}{\partial \zeta}\right)\nn
%\end{eqnarray}
%This form although elegant has been proven to be unappropriate for a numerical evolution.

The perturbation equations \eqref{eq:explicit_v2} are complemented by boundary conditions which describe the behaviour of the perturbations on the boundaries of the numerical domain, which are the rotation axis and the surface of the star. One also has to discriminate between perturbation variables with scalar and vectorial character to find the correct conditions along the rotation axis. In our case, the energy density perturbation which is described by $H$ and the $\zeta$-component of the perturbed 4-velocity $\delta u_{\zeta}$, described by $f_{2}$ are clearly the scalar perturbations while $f_{1}$ and $f_{3}$ (describing velocity perturbations in a $\zeta = {\textrm const.}$-plane) are the vectorial perturbations.
Let us first look at the boundary conditions at the surface, since this is very easy in our formulation. Since the perturbation functions on the lefthand side of \eqref{eq:perturbedAnsatz} drop to zero there, all our perturbation quantities vanish. So on the surface of the neutron star we have
\begin{equation}
\label{eq:boundaryCondSurface}
f_{1}|_{\rm{surface}} = f_{2}|_{\rm{surface}} = f_{3}|_{\rm{surface}} = H|_{\rm{surface}} = 0\,. 
\end{equation}
For the rotation axis we have to consider three different cases, depending on the value of $m$ and the nature of the perturbation variable. Scalar perturbations have to be unique along the axis for all values of $m$ when varying the azimuthal angle $\varphi$. Vectorial perturbations have to change like $\cos(\varphi)$ or $\sin(\varphi)$ when varying $\varphi$. This means, that only for $m=\pm 1$ they are allowed to have nonzero values along the axis. %\comment{I am not extremely happy with the last two sentences} 
If we then take into account again the equations \eqref{eq:limes} and \eqref{eq:perturbedAnsatz}, we end up with the boundary conditions for the rotational axis depicted in table \ref{tb:axisCond}.
%%%%
%\pagebreak
\begin{table}[h!]
\begin{tabular}{| c | c | c | c | c |}
\hline
$~~m$-value~~ & $~~f_{1}|_{\rm{axis}}$~~ & ~~$f_{2}|_{\rm{axis}}$~~ & ~~$f_{3}|_{\rm{axis}}$~~ & $ ~~H|_{\rm{axis}}$~~ \\
\hline
0 & 0 & 0 & 0 & ~~finite \& continuous~~ \\
\hline
$\pm 1$ & 0 & 0 & ~~finite \& continuous~~ & 0 \\
\hline
else & 0 & 0 & 0 & 0\\
\hline
\end{tabular}
\caption{Boundary condition for the perturbation variables along the rotational axis}
\label{tb:axisCond}
\end{table}
\noindent

%%%%%%%%%%%%%%%%%%%%%%%%%  SEC. III: DESCRIPTION OF NUMERICAL METHODS  %%%%%%%%%%%%%%%%%%%%%%%%%%%%%%%

\section{Numerical method}
\label{sec:numMethod}
As described already in the previous section, in this study we adopted a comoving frame of reference, in which the metric takes the form shown in equation \eqref{eq:lineElement}. The numerical method used to solve the equations governing the stellar background for this special metric is described in detail in \cite{Ansorg:2003mz}; here we will briefly summarize the parts which are crucial for our work. Since the stationary background model possesses rotational symmetry as well as symmetry with respect to the equatorial plane, for the computation it is sufficient to consider the physical domain
\begin{equation}
\label{eq:physDomain}
\mathcal{D}_{+} = [(\rho, \zeta)\,,\rho\geq 0, \zeta \geq 0]\,.
\end{equation}
By means of a coordinate transformation
\begin{equation}
\label{eq:coordTransf}
\mathcal{T}_{+} = [(\sigma,\tau)\in [0,1]\times [0,1]\,,(\rho = \rho(\sigma,\tau), \zeta = \zeta(\sigma,\tau)) \in \mathcal{D}_{+}]
\end{equation}
the neutron star interior is mapped onto the unit square and then, the equations are solved in this new coordinate system with a spectral-methods--code.

\begin{figure}[htp!]
\centering
\includegraphics[width=3.2in]{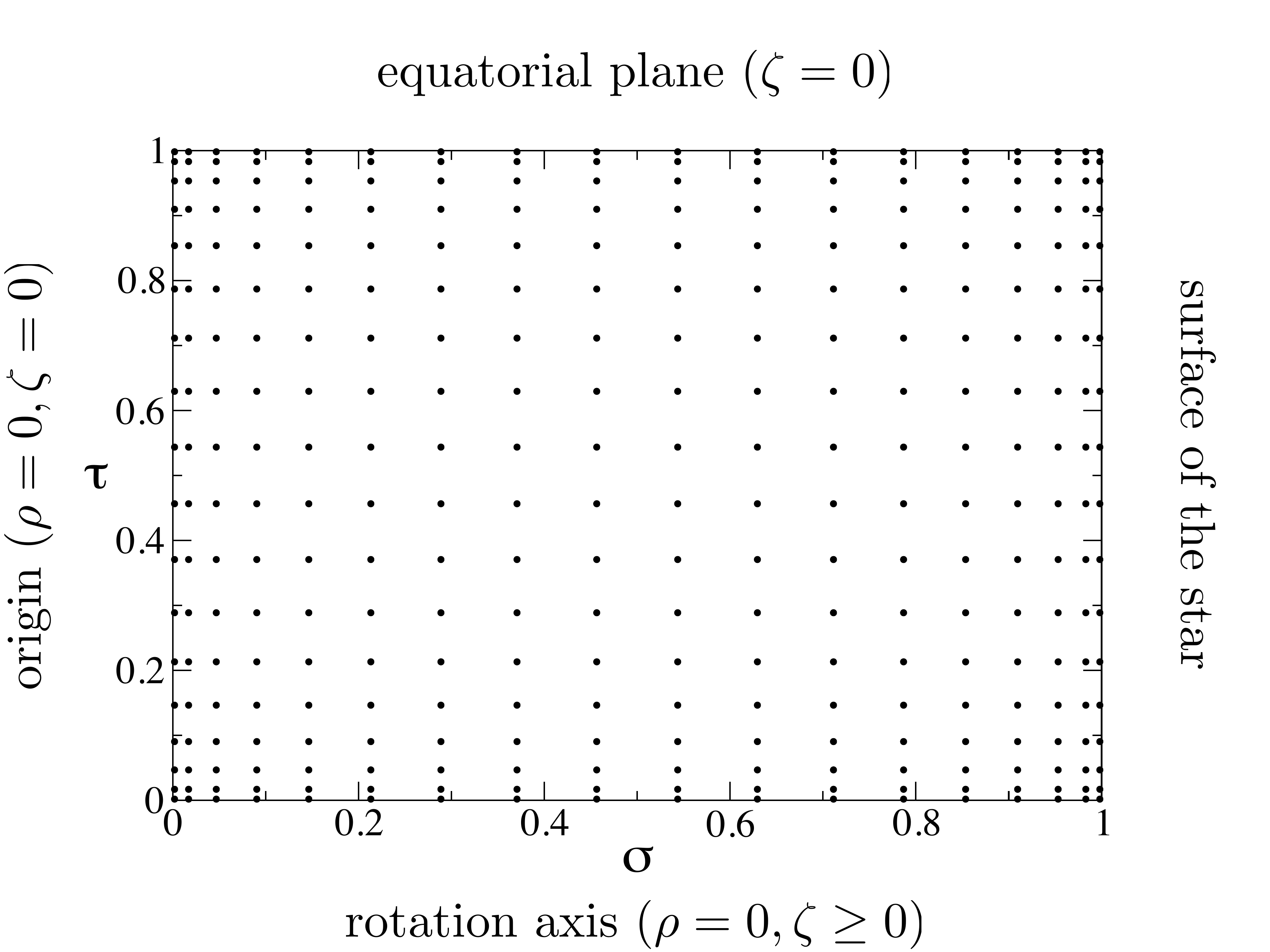}
\caption{Layout of the numerical grid used in constructing the stationary neutron star background model}
\label{fig:akmGrid}
\end{figure}

\begin{figure}[h!]
\centering
\includegraphics[width=3.2in]{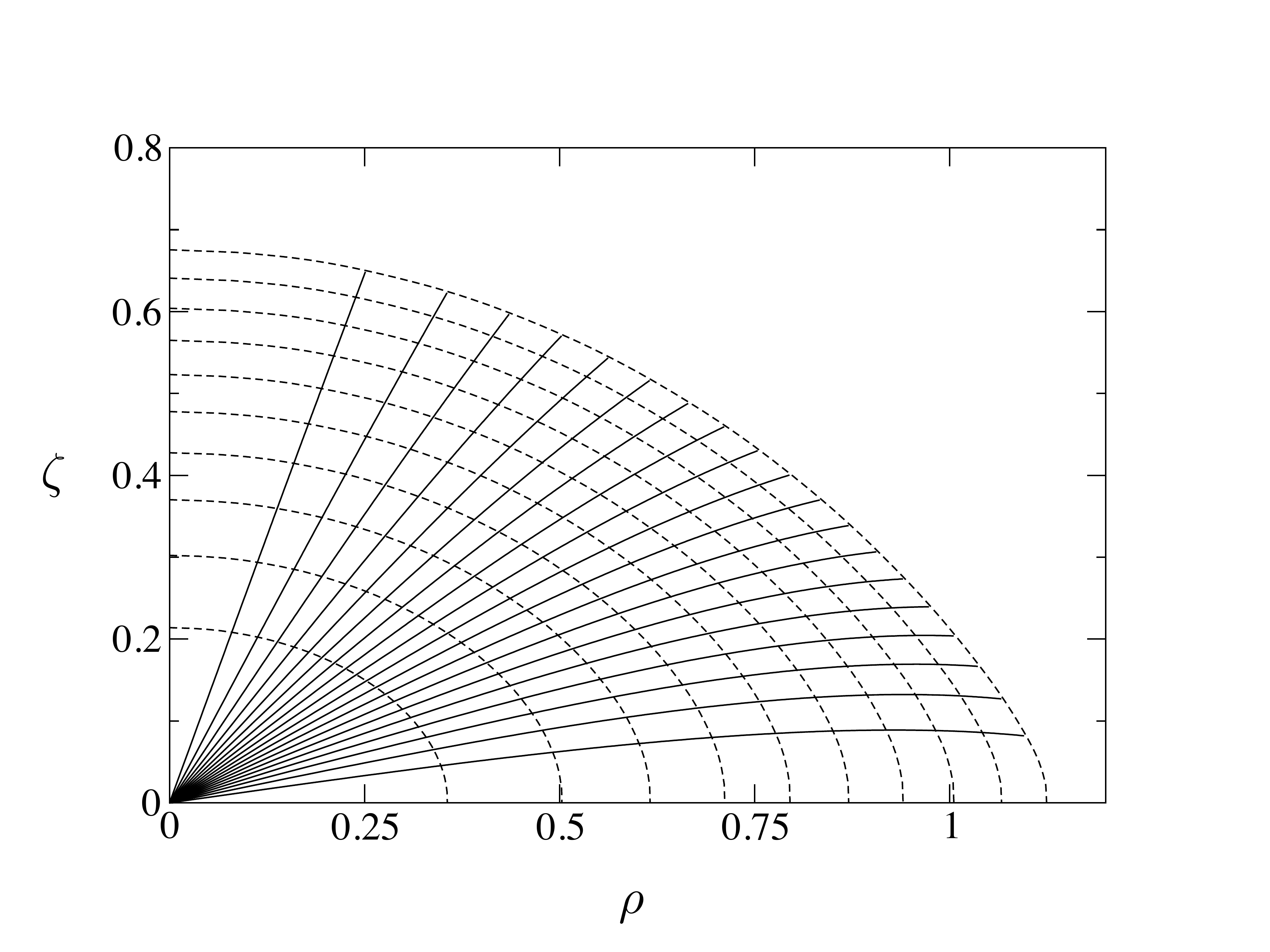}
\caption{Some $\tau=const.$ (solid) and $\sigma=const.$ (dotted) coordinate lines in the $(\rho, \zeta)$ coordinate system}
\label{fig:akmGridLines}
\end{figure}

For  example,  the distribution of the grid points is shown in Figure \ref{fig:akmGrid} for a resolution of $18\times 18$. Note, that the special choice of collocation points guarantees that there will never be any grid points directly on the boundaries. One can also observe the typical clustering of the grid points near the boundary which is characteristic for spectral methods.
Also in this figure, some properties of the coordinate transformation $\mathcal{T}_{+}$ are labelled. One can notice, that the surface of the neutron star gets mapped onto $\sigma=1$; this is independent of the rotation rate, even for rapidly rotating neutron stars this coordinate line always corresponds to the stellar surface. Additionally, $\tau=0$ corresponds to the rotation axis above the equatorial plane and $\tau=1$ to the equatorial plane itself. As mentioned earlier, since the background star is axisymmetric as well as symmetric with respect to the equatorial plane, this computational domain suffices in order to construct the background model. Note also, since we have four boundaries in our computational domain, but only three ``physical" boundaries (i.e. rotation axis, equatorial plane and stellar surface), one single point gets smeared into a coordinate line; in our case, $\sigma=0$ corresponds to the center of the star (i.e. $\rho = \zeta = 0$).
\noindent
The coordinates $(\sigma,\tau)$ are similar to their spherical coordinates counterparts $(r,\theta)$ in the sense that moving from $\tau=0$ to $\tau=1$ on an arbitrary $\sigma=const.$-line means to start from a point at the rotation axis and move somewhat ``parallel" to the surface to a point at the equatorial plane. Vice versa if one moves along an arbitrary $\tau=const.$-line then one starts from the center of the star to the surface. Figure \ref{fig:akmGridLines} illustrates this for a rapidly rotating stellar model with a ratio of polar coordinate radius to equatorial coordinate radius of $r_{p}/r_{e}= 0.6$.
\noindent
This compact star is rotating near its Kepler-limit and obviously is non-spherical. Nevertheless, as discussed above, the surface of the stellar model is described by $\mathcal{S}= [(\rho(1,\tau), \zeta(1,\tau))\,, \tau\in[0,1]]$.

For the linear perturbations under study these coordinates suffice and we used it with minor modifications. Because of the equatorial symmetry of the background it is sufficient to use $\mathcal{D}_{+}$ (see \eqref{eq:physDomain}) as the domain of computation, but this is no longer true for general axisymmetric perturbations. There are two ways to circumvent this. The first is based on the fact, that every perturbation can be decomposed into a symmetric and into an anti-symmetric part. To illustrate this, consider $\mathcal{A}_{+}= [x\,,x\geq 0]$ as a one-dimensional computational domain and $f(x,t)$ as an arbitrary perturbation with values in $\mathcal{A}_{-}= [x\,,x<0]$ too. If $x=0$ corresponds to the axis of symmetry we have
\begin{equation*}
f(x,t)= f_{s}(x,t) + f_{a}(x,t)
\end{equation*}
with
\begin{equation*}
f_{s}(x,t) = \frac{f(x,t) + f(-x,t)}{2}\hspace{0.2in}{\rm and}\hspace{0.2in}f_{a}(x,t) = \frac{f(x,t) - f(-x,t)}{2}\,.
\end{equation*}
The procedure for a time-evolution of an arbitrary initial perturbation $f(x, t_{0})$ on $\mathcal{A}:=\mathcal{A}_{+}+\mathcal{A}_{-}$ would be to decompose any initial perturbation into its symmetric and anti-symmetric parts, perform an evolution of these parts, complement the solutions according to their symmetry behaviour for $x<0$ and add the solutions up. Note that in this case there are two different boundary conditions to impose on the unknown function at $x=0$. For the symmetric  part we have $\partial_{x}f_{s}(x,t)|_{x=0} = 0$, for the antisymmetric part we have $f_{s}(x,t)|_{x=0} = 0$. Note also, that in order to study eigenmodes and compute their eigenfrequencies it is sufficient to restrict these studies on either symmetric or anti-symmetric perturbations; for this purpose it is not neccessary to consider arbitrary initial data.

Since the goal was to create a code that can handle arbitrary perturbations, we chose the second option in handling this problem, that is we extend our physical domain to include also the region
\begin{equation}
\label{eq:addDomain}
\mathcal{D}_{-} = [(\rho, \zeta)\,,\rho\geq 0, \zeta < 0]\,.
\end{equation}
In practice we ``glue" two copies of $\mathcal{D}_{+}$ together along the equatorial plane. Correspondingly the computational domain extents in the $\tau$-dimension beyond $\tau = 1$, i.e. the $\theta$-direction in analogy to spherical coordinates we discussed earlier. Hence the extended domain is given by
\begin{equation}
\label{eq:newCoordTransf}
\mathcal{T} = [(\sigma,\tau)\in [0,1]\times [0,2]\,,(\rho = \rho(\sigma,\tau), \zeta = \zeta(\sigma,\tau)) \in \mathcal{D}:= (\mathcal{D}_{+} + \mathcal{D}_{-})]
\end{equation}
With this choice, there are no other boundary conditions to impose than those described in section \ref{sec:pertEq}; the equatorial plane now lies in the interior of the computational domain and its boundaries are given by the center ($\sigma=0$), the surface ($\sigma=1$), the part of the rotation axis above the equatorial plane ($\tau=0$) and the corresponding part underneath the equatorial plane ($\tau=2$). Of course one pays a price for studying arbitrary perturbations: This grid is now twice as large as before; this means for a 2D-code an increase in computation time by a factor of 4.

Special care has to be taken for the correct transformation behaviour of our various background quantitites which, up to now, are only known in $\mathcal{D}_{+}$. In addition to simply ``mirror" all neccessary quantities along the equatorial plane (i.e. along $\tau=1$), some derivatives need to be multiplied by a factor of $-1$. Derivatives with respect to $\rho$ pose no problem; if $(\sigma,\tau)$ denotes a grid point in $\mathcal{D}_{+}$ (i.e. $\tau\leq 1$) and $g$ is an arbitrary background quantity, we have
\begin{equation}
\label{eq:behaviourOfRhoDerivs}
\frac{\partial g}{\partial\rho}|_{(\sigma,\tau)} = \frac{\partial g}{\partial\rho}|_{(\sigma,2-\tau)}\,.
\end{equation}
This is no longer true when one uses derivatives with respect to $\zeta$; in this case it is
\begin{equation}
\label{eq:behaviourOfZetaDerivs}
\frac{\partial g}{\partial\zeta}|_{(\sigma,\tau)} = -\frac{\partial g}{\partial\zeta}|_{(\sigma,2-\tau)}\,.
\end{equation}
Yet another transformation behaviour has to do with the fact, that the equations we study (i.e. system \eqref{eq:explicit_v2}) are written in $(\rho,\zeta)$-coordinates but we use as numerical domain our extended $(\sigma,\tau)$-system. For our perturbation variables we have to switch between these systems. So in addition to all background quantities and their derivatives with respect to $\rho$ and $\zeta$, the transformations coefficients $\partial \sigma/\partial\rho$, $\partial \sigma/\partial\zeta$, $\partial \tau/\partial\rho$ and $\partial \tau/\partial\zeta$ are also available in $\mathcal{D}_{+}$. With these coefficients it is possible to compute for any perturbation quantitiy $f$ given on the numerical domain $\mathcal{T}$ the values of
\begin{eqnarray}
\label{eq:gridVariableDeriv}
\frac{\partial f}{\partial\rho} &=& \frac{\partial f}{\partial \sigma}\frac{\partial \sigma}{\partial\rho} + \frac{\partial f}{\partial \tau}\frac{\partial \tau}{\partial\rho}\\
\frac{\partial f}{\partial\zeta} &=& \frac{\partial f}{\partial \sigma}\frac{\partial \sigma}{\partial\zeta} + \frac{\partial f}{\partial \tau}\frac{\partial \tau}{\partial\zeta}\nn
\end{eqnarray}
we need to know for the righthand-sides of the evolution equations \eqref{eq:explicit_v2}.
Equations \eqref{eq:gridVariableDeriv} are obviously  valid for the background quantities as well and this helps in finding the correct form of the transformation coefficients when going from $\mathcal{D}_{+}$ to $\mathcal{D}_{-}$. We know how the $\rho$- and $\zeta$-derivatives (i.e. the lefthand-sides of \eqref{eq:gridVariableDeriv}) transform in $\mathcal{D}_{-}$ as well what happens to their $\sigma$- and $\tau$-derivatives there. This leads to
\begin{eqnarray}
\label{eq:behaviourOfTransfCoeff_1}
\frac{\partial \sigma}{\partial\rho}|_{(\sigma,\tau)} & = &  \frac{\partial \sigma}{\partial\rho}|_{(\sigma,2-\tau)}\quad \mbox{and} \quad
\frac{\partial \sigma}{\partial\zeta}|_{(\sigma,\tau)}  =  -\frac{\partial \sigma}{\partial\zeta}|_{(\sigma,2-\tau)}
\end{eqnarray}
and
\begin{eqnarray}
\label{eq:behaviourOfTransfCoeff_2}
\frac{\partial \tau}{\partial\rho}|_{(\sigma,\tau)} & = & -\frac{\partial \tau}{\partial\rho}|_{(\sigma,2-\tau)}\quad \mbox{and} \quad
\frac{\partial \tau}{\partial\zeta}|_{(\sigma,\tau)}  =  \frac{\partial \tau}{\partial\zeta}|_{(\sigma,2-\tau)}
\end{eqnarray}
We then use standard finite-differencing schemes and time integrators to solve system \eqref{eq:explicit_v2}. However, the numerical evolution of the equations was unstable i.e. after the  first few time steps, high frequency oscillations of exponentially growing modes developed near the center of the star (i.e. at $\sigma=0$). This is most likely due to the presence of a coordinate singularity at the origin but also because some of the coefficients on the righthand-sides of our evolution system get nearly singular when moving to the center (compare the denominators of \eqref{eq:explicit_v2} with \eqref{eq:Fdefinition} and \eqref{eq:limes}). However, the occurrence of singular terms in general doesn't neccessarily lead to a failure of the numerical scheme.

The solution to this problem was the utilization of additional viscous terms in the evolution equations. Since they do not represent any physical effect in the initial setup of the problem, they are commonly referred to as {\it artificial viscosity}. Here, to each of the four equations  we added a Kreiss-Oliger like term of the form (see  \cite{Gustafsson:1995yq} for details)
\begin{equation}
\label{eq:artViscosity}
\mathcal{V}(f) = \alpha\left(D^{+}_{\sigma}D^{-}_{\sigma} + D^{+}_{\tau}D^{-}_{\tau} \right)f\,,
\end{equation}
where $f$ is a perturbation variable, $\alpha = const.$ is the dissipation coefficient, $(\sigma,\tau)$ are the spatial coordinates and $D^{+}$, $D^{-}$ are the standard forward- and backward--difference operators. By using dissipation coefficients $\alpha_{i}\,, i = 1\ldots 4$ ranging from $10^{-3}-10^{-4}$ it become possible  to stabilize the time-evolution code against exponentially growing instabilities.

As already described earlier in section \ref{sec:pertEq}, after a successfull simulation, the real part of the complex solution is taken and a discrete Fourier-transformation at several points inside the star is performed on these data to extract the oscillation frequencies. If $N$ is the number of points in this time series and $\Delta t$ is the temporal resolution, the corresponding frequency resolution and the Nyquist-frequency (i.e. the maximum frequency that can be resolved) are given by
\begin{equation}
\label{eq:frequencyStuff}
\Delta f = \frac{1}{N\Delta t} \quad{\textrm {and}}\quad f_{c} = \frac{1}{2\Delta t}
\end{equation}
Since an explicit numerical scheme has been used for time-evolution, there are certain restrictions on the absolute value of $\Delta t$. The timestep cannot be arbitrarily large and depends on the spatial resolution of our grid (CFL-criterion). For most of the simulations,  a timestep of the order of $\Delta t\approx 10^{-6}$ sec and $N\approx 10^{4}$ was used. The total evolution time then is $t_{max}\approx 30-40$ ms with $\Delta f \approx 15-30$ Hz and $f_{c}\approx 8 - 12$ kHz; details are following in the next section \ref{sec:results}.

%%%%%%%%%%%%%%%%%%%%%%%%%  SEC. IV: RESULTS %%%%%%%%%%%%%%%%%%%%%%%%%%%%%%%

\section{Results}
\label{sec:results}
For a spherical symmetric background and even in the slow-rotation approximation, the angular dependence of a mode is often described in terms of spherical harmonics $Y_{lm}$. If the perturbation changes like $(-1)^{l}$ when applying the transformation $\mathbf{r}\rightarrow -\mathbf{r}$ it is called an even or polar mode while odd or axial modes behave like $(-1)^{l+1}$. In the axisymmetric case (i.e. for $m=0$) all these modes are stable while non-axisymmetric perturbations may become unstable to the CFS-instability \cite{1970PhRvL..24..611C,1978ApJ...222..281F}.

\subsection{Axisymmetric case}
\label{ssec:Axi}
Axisymmetric perturbations, due to their simplicity have been studied in detail with perturbative methods but mainly via evolution of non-linear equations.
\subsubsection{Polar perturbations}
\label{sssec:axiAndPolar}
The results of this approach will be compared with the non-linear results published in \cite{Font:2001eu}. There, a relativistic hydrodynamics code is used to study stellar oscillations in the Cowling approximation. The background models are commonly referred to as BU; they are uniformly rotating neutron stars with a polytropic equation of state with $\Gamma = 2$, $K = 100$ and fixed central rest-mass density $\rho_{c} = 1.28\times 10^{-3}$ in units where $G = c = M_{\odot} = 1$. In the nonrotating case, this leads to a stellar model with a gravitational mass of $M_{0} = 1.4\,M_{\odot}$ and a circumferential radius of $R = 14.15$ km. The applied initial perturbations were decomposed into spherical harmonics and for $l = 0$ a density perturbation of the form
\begin{equation}
\label{eq:lEqualsZeroPert}
\delta\rho = A\rho_{c}\sin\left(\frac{\pi r}{r_{s}(\theta)}\right)
\end{equation}
is used as initial data. Here $A$ is the perturbation amplitude, $(r,\theta)$ denote spherical coordinates and $r_{s}(\theta)$ is the coordinate radius of the star (which is not independent of $\theta$ when the star is rotating) in this spherical system. In our simulations we mainly use a $18\times 18$ spectral grid to compute the background (this already gives a very accurate stellar model with an accuracy of $10^{-10}$) and interpolate to our computatonal domain, which can have practically any arbitrary resolution; mainly we use $50\times 40$, $100\times 80$ or $200\times 160$ grid points.

The following figure \ref{fig:l0TimeSeriesAndTransform} shows in the left panel a 12 msec-long section of a simulation with initial data described by \eqref{eq:lEqualsZeroPert} and a nonrotating background model on a $50\times 40$ grid. The right panel depicts the logarithm of the one-sided power spectral density of the complete time series with a frequency resolution of $\Delta f = 20$ Hz; the data for this figure were extracted from the spatial position $(\sigma,\tau) = (0.5, 0.5)$.\\
\begin{figure}[ht!] 
\centering
\includegraphics[width=0.465\textwidth]{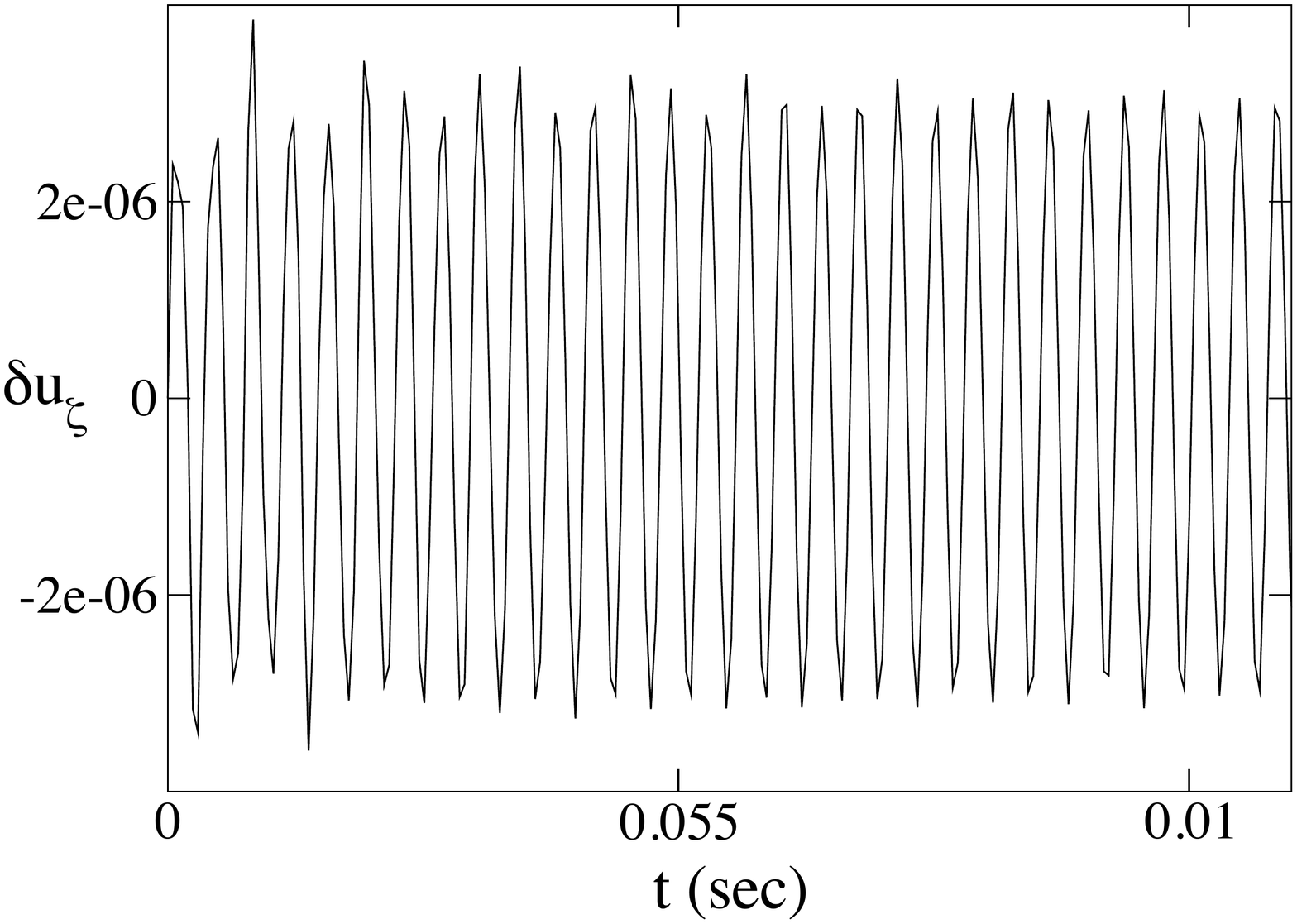}
\hspace{0.1in}
\includegraphics[width=0.47\textwidth]{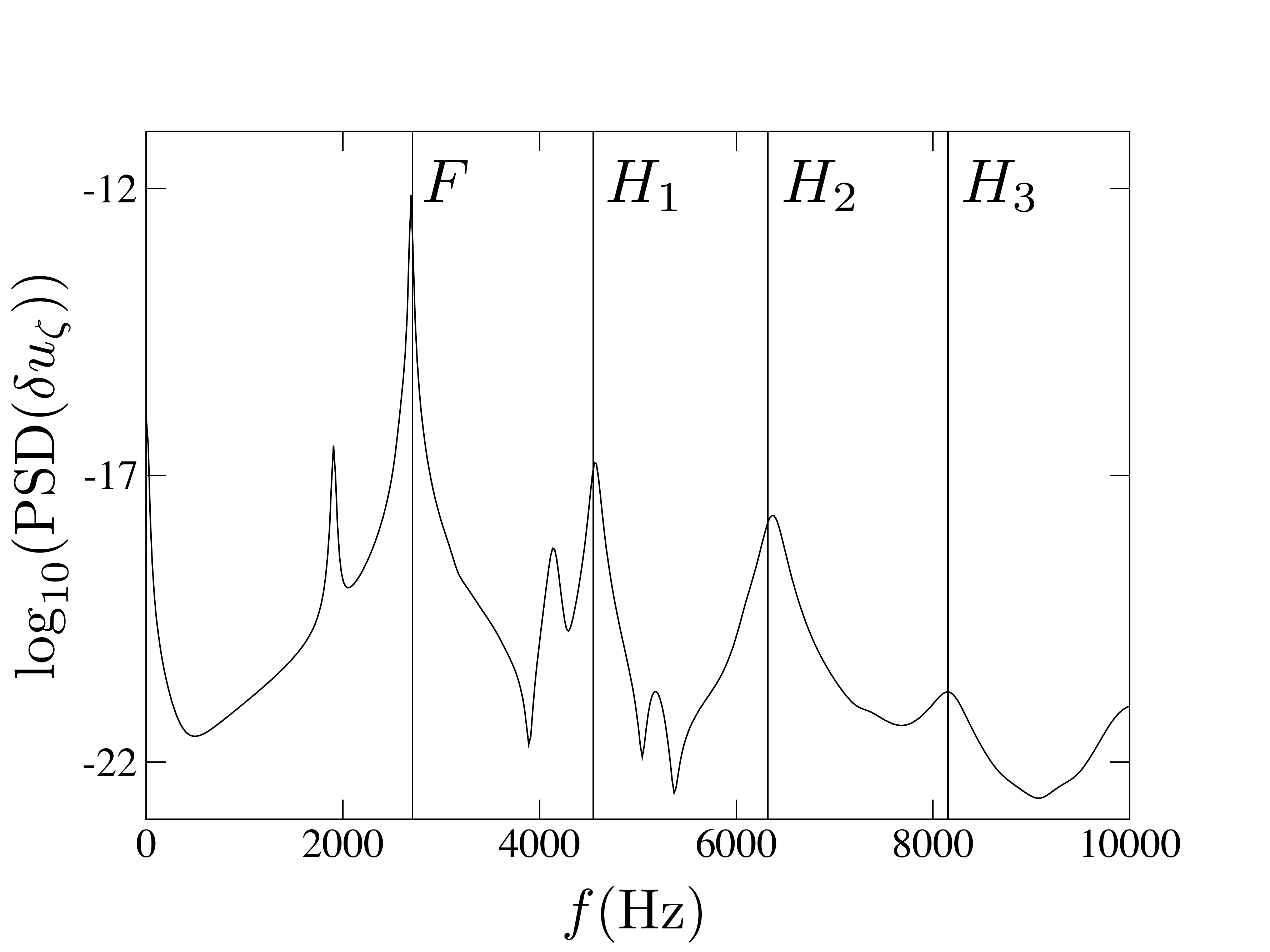}
\caption{{\it Left panel:} Time-evolution of the $\zeta$-component of the perturbed 4-velocity for an axisymmetric pressure perturbation {\it Right panel:} The corresponding Fourier transform of the $\delta u_{\zeta}$- time series. The fundamental radial mode together with a few overtones are apparent}
\label{fig:l0TimeSeriesAndTransform}
\end{figure}
\noindent
In the frequency plot one can see several peaks, some of them are already labelled; the vertical lines are the equivalent frequencies one can find in \cite{Font:2001eu}. The strongest and one of the sharpest peaks is the one at $f = 2.705$ kHz and belongs to the fundamental quasi-radial oscillation mode ($F\,$-mode) with no nodes of the corresponding eigenfunction in the radial direction. Alongside with this oscillation some other modes were excited as well, most notably several overtones of the F-mode labelled $H_{1}$, $H_{2}$ and $H_{3}$ which have one, two or three nodes of their eigenfunctions in the radial direction.

Additionally, to the estimation of the mode frequencies from a given time evolution via a Fourier transform we implemented a method to retrieve their corresponding eigenfunctions. The amplitude of the eigenfunction at a given point directly correlates to the strength of the corresponding peak in the power spectral density at this particular point. In order to extract the eigenfunction of a specific oscillation mode one has to iterate over the computational domain, taking Fourier transforms at many grid points and monitor the variation in amplitude of the mode peak one wants to study. This eigenfunction can then be used as an improvement to the first trial eigenfunction and can be put back as initial data for another simulation. Usually this procedure which is called {\it mode recycling}, when applied repeatedly, will enhance and sharpen the mode that is recycled and will suppress additional modes that are excited.

\begin{figure}[ht!]
\centering
\includegraphics[width=4.7in]{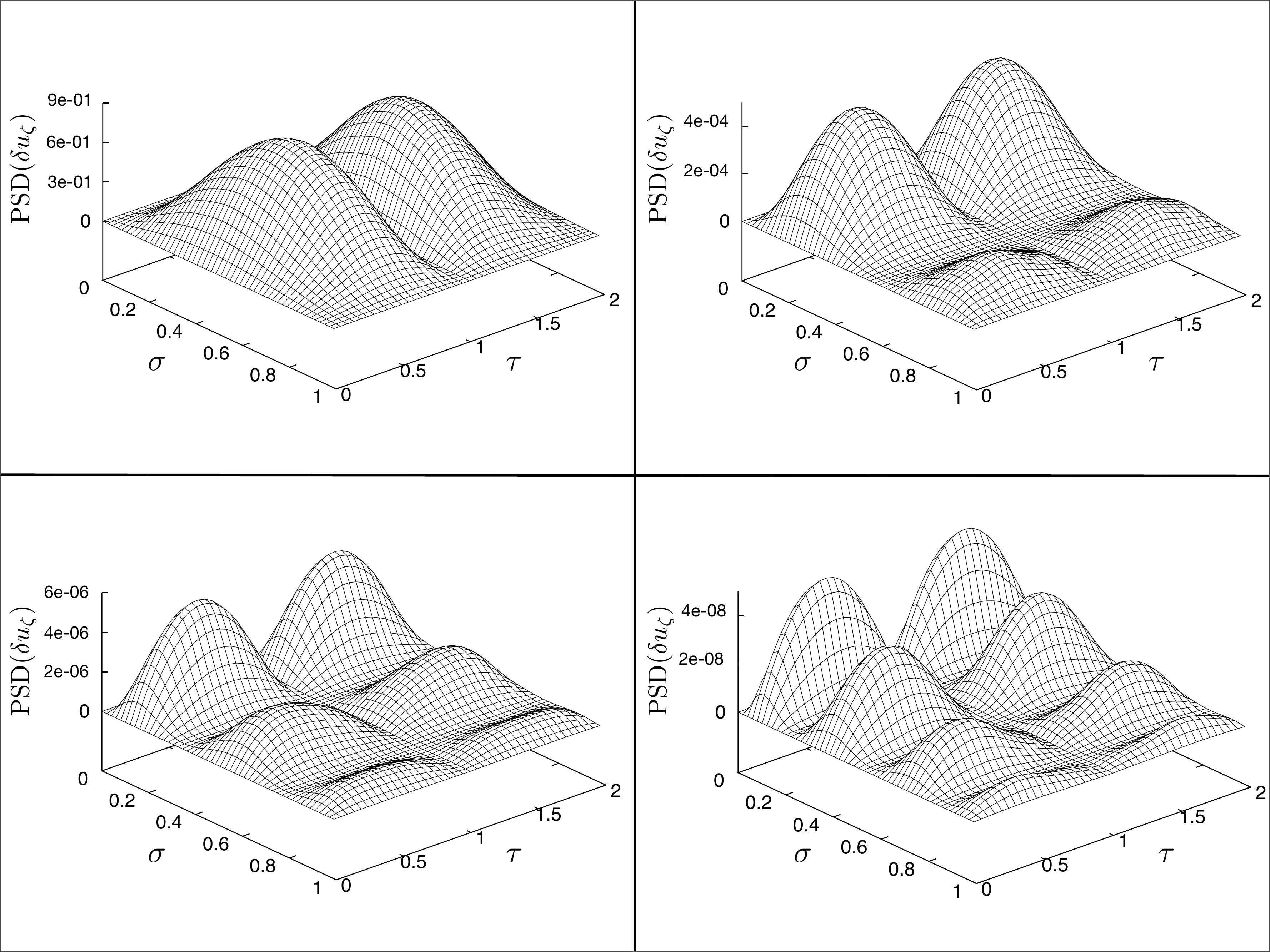} 
\caption{{\it Top Row:} The amplitude of the eigenfunctions for the $F$- and $H_{1}\,$-mode {\it Bottom Row:} The amplitude of the eigenfuctions for the $H_{2}$- and $H_{3}\,$-mode }
\label{fig:eigenfunctions1}
\end{figure}

In figure \ref{fig:eigenfunctions1} the absolute values of the eigenfunctions corresponding to the oscillation modes labelled in the previous figure were extracted using this technique. As already discussed earlier it is $(\sigma,\tau)\in [0,1]\times[0,2]$ and $\tau=1$ describes the equatorial plane.

\noindent
One can see, that $\delta u_{\zeta}=0$ along the equatorial plane and also that the number of radial nodes increases along the sequence $F, H_{1}, H_{2}, H_{3}$.

The second strongest peak in figure \ref{fig:l0TimeSeriesAndTransform} at $f=1.929$ kHz (it is already roughly two orders of magnitude lower than the $F\,$-mode peak) does not belong to a quasi-radial oscillation. Instead, the extracted eigenfunction shows an angular dependence that is in agreement with an axisymmetric quadrupolar perturbation. By extracting the eigenfunction for the pressure perturbation of this mode one gets a useful initial perturbation for extracting non-radial modes and eigenfunctions. The result of such a simulation is depicted in the next figure \ref{fig:spectrumAndEigenRecycled}.

\begin{figure}[ht!] 
\centering
\includegraphics[width=0.47\textwidth]{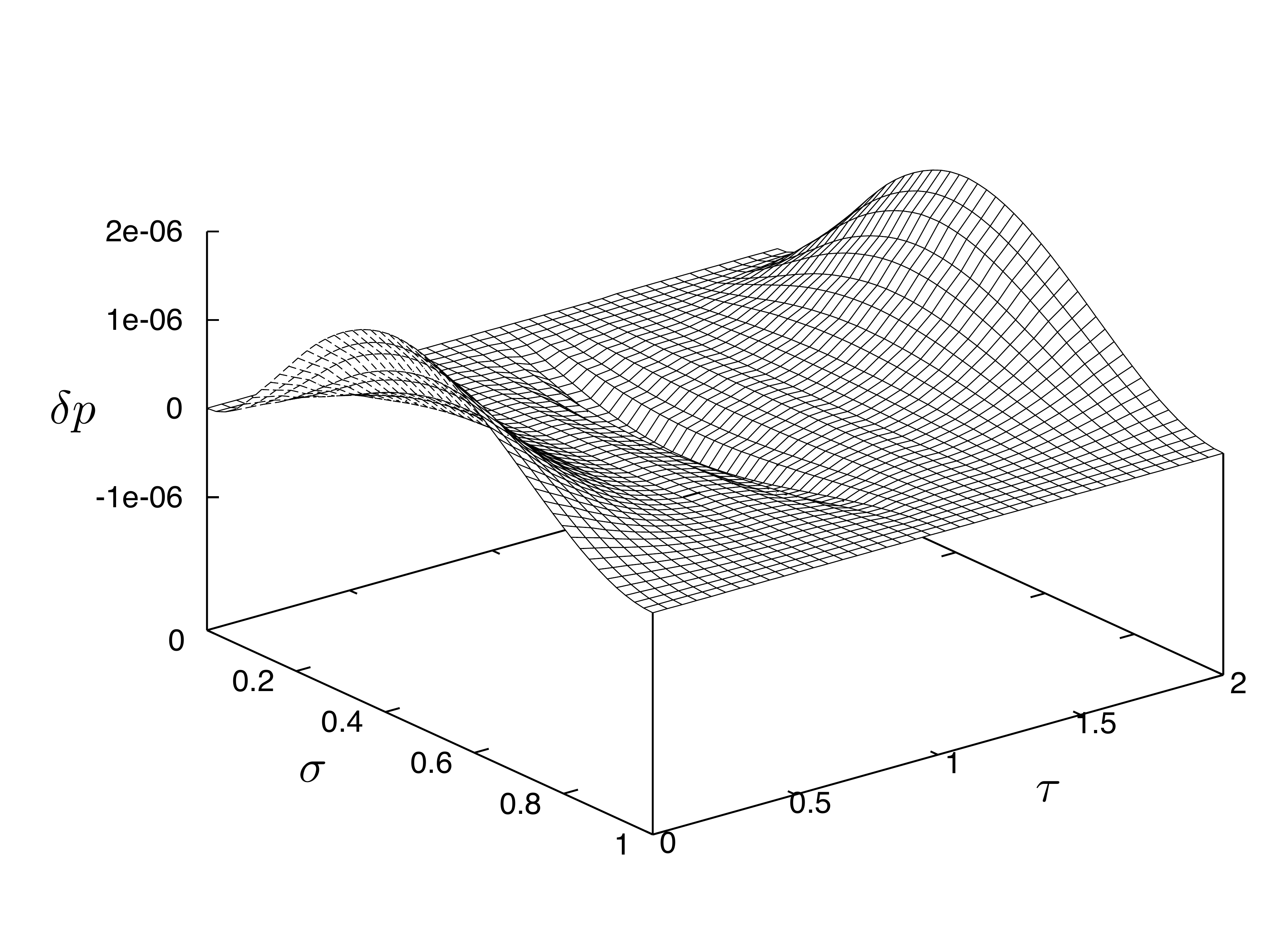}
\hspace{0.1in}
\includegraphics[width=0.46\textwidth]{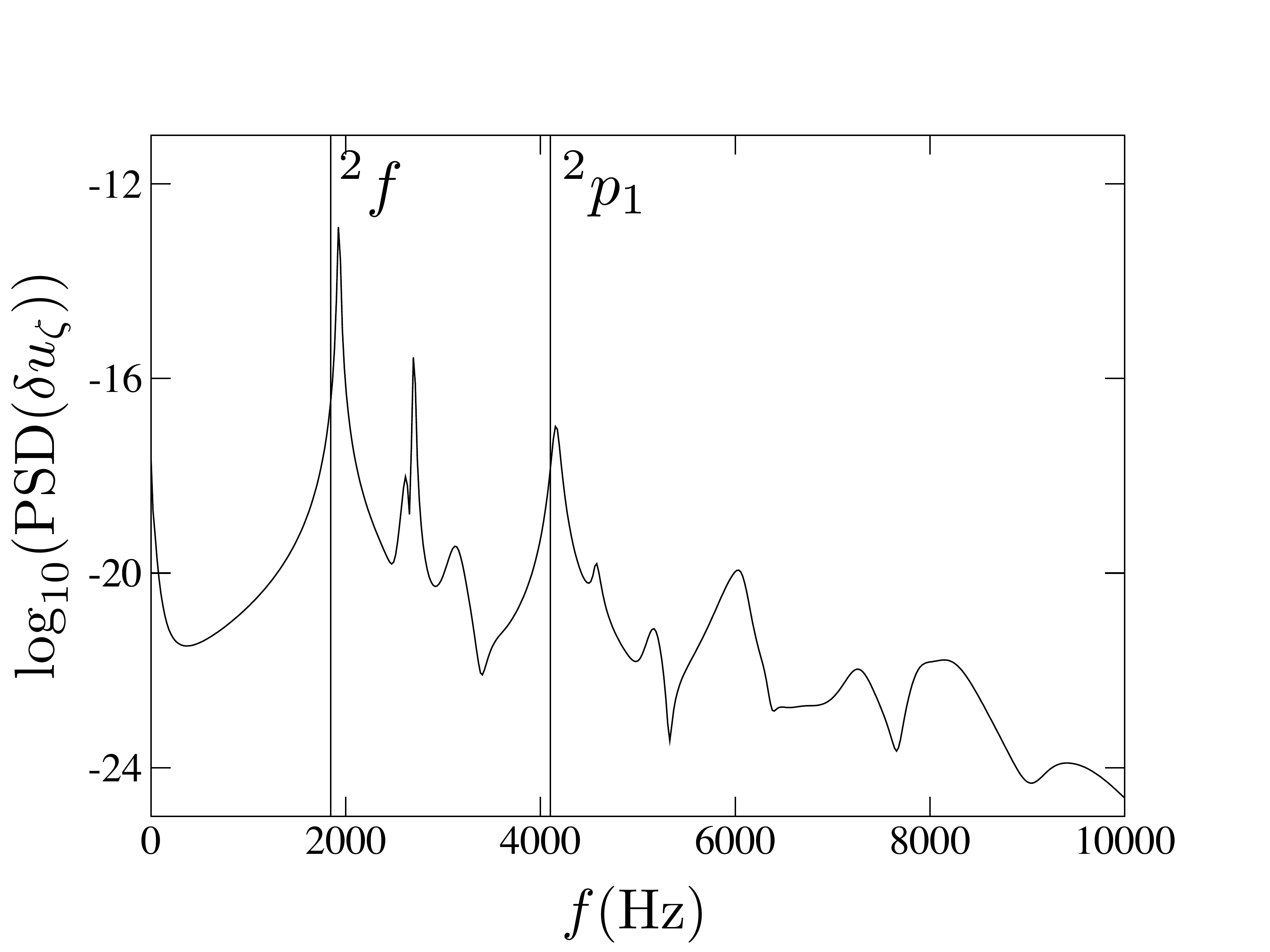}
\caption{The shape of the recovered pressure eigenfunction used for a mode recycling run and Fourier transform of the $u_{\zeta}$- time series with the pressure eigenfunction as initial data}
\label{fig:spectrumAndEigenRecycled}
\end{figure}
As one can easily notice, the $^{2}f\,$-mode is the dominant oscillation while the first $(l = 2, m = 0)$-overtone, the $^{2}p_{1}\,$-mode, has been also considerably enhanced. In this simulation, both modes are stronger than their $l=0$ counterparts and several other modes which were not clearly visible in the first run (compare to figure \ref{fig:l0TimeSeriesAndTransform}) become noticable. The corresponding values taken from \cite{Font:2001eu} for comparison are indicated by solid vertical lines in the power spectral density plot. Here we notice larger deviations than in the previous radial case but they are smaller than 5\%.
%%%
\begin{table}[ht!]
\begin{tabular}{| c | c | c | c | c |}
\hline
~~$\Omega$ (kHz)~~ & ~~$F$ (kHz)~~ & ~~$H_{1}$ (kHz)~~ & ~~$H_{2}$ (kHz)~~ & ~~$H_{3}$ (kHz)~~ \\
\hline
0.0 		& 2.679	& 4.561	& 6.380	& 8.178\\
2.182   	& 2.638	& 4.466	& 6.255	& 8.035\\
3.062   	& 2.605	& 4.435	& 6.253	& 8.058\\
3.712   	& 2.570	& 4.409	& 6.275	& 8.111\\
4.229   	& 2.539	& 4.410	& 6.310	& 8.156\\
4.647   	& 2.500	& 4.400	& 6.330	& 8.237\\
4.976   	& 2.484	& 4.392	& 6.356	& 8.334\\
5.213   	& 2.456	& 4.390	& 6.370	& 8.405\\
5.344   	& 2.426	& 4.394	& 6.380	& 8.411\\
\hline
\end{tabular}
\caption{Frequencies of the axisymmetric modes $F$, $H_{1}$, $H_{2}$ and $H_{3}$ for the BU model at different rotation rates}
\label{tb:BUmodelAxisymmetric1}
\end{table}

One can increase the rotation rate of the BU model and estimate how the various frequencies will change when the neutron star rotates faster and faster. We did this calculation for three resolutions starting from $50\times 40$ and doubling it twice. The results for the already discussed $l=0\,$-modes at the highest resolution (i.e. $200\times 160$) and with $\Delta f = 20$ Hz are summarized in Table \ref{tb:BUmodelAxisymmetric1}.

\begin{table}[ht!]
\begin{tabular}{| c | c | c  |}
\hline
~~$\Omega$ (kHz)~~ & ~~$^{2}f$ (kHz)~~ & ~~$^{2}p_{1}$ (kHz)~~ \\
\hline
0.0     	& 1.890	& 4.130\\
2.182   	& 1.890	& 4.065\\
3.062   	& 1.906	& 3.970\\
3.712   	& 1.895	& 3.838\\
4.229   	& 1.875	& 3.674\\
4.647   	& 1.844	& 3.487\\
4.976   	& 1.794	& 3.275\\
5.213   	& 1.703	& 3.056\\
5.344   	& 1.613	& 2.426\\
\hline
\end{tabular}
\caption{Frequencies of the axisymmetric quadrupolar modes $^{2}f$ and $^{2}p_{1}$ for the BU model at different rotation rates}
\label{tb:BUmodelAxisymmetric2}
\end{table}
%\noindent%\noindent
In Table \ref{tb:BUmodelAxisymmetric2} the oscillation frequencies of the $^{2}f$- and the $^{2}p_{1}$-mode are shown for various rotation rates.
In general, these results are in good agreement with published values; their absolute differences never exceed the 5\%-level. By increasing the resolution the most significant improvement has been observed when doubling the low $50\times 40\,$-resolution. For some modes the change of the frequency in these two resolutions is rather dramatic compared to the change when improving from the medium to the high resolution. We take this as an indication that for most purposes a resolution of $100\times 80$ grid points suffices to get already quite accurate results.
We also did convergence checks on the two strongest modes for $l=0$ and $l=2$ at our three basic resolutions and an additional one with $75\times 60$ gridpoints. The Iterated Crank-Nicholson scheme, which is used for time-evolution here, is first order accurate in time and second order accurate in space. By increasing the number of gridpoints one expects to observe a quadratic convergence in the perturbation functions and this was indeed the case.\\
The following figure \ref{fig:resolutionDependence} shows how the frequencies of the $F$-, $H_{1}$-, $^{2}f$- and $^{2}p_{1}\,$-mode change as functions of the rotation rate and resolution; there the line connecting circles are the values taken from \cite{Font:2001eu}. Especially the frequencies for the first $p$-mode corresponding to $l=2$ agree very well with them; also the other modes show a similar behaviour.
 
\begin{figure}[htp!]
\centering
\includegraphics[width=4.7in]{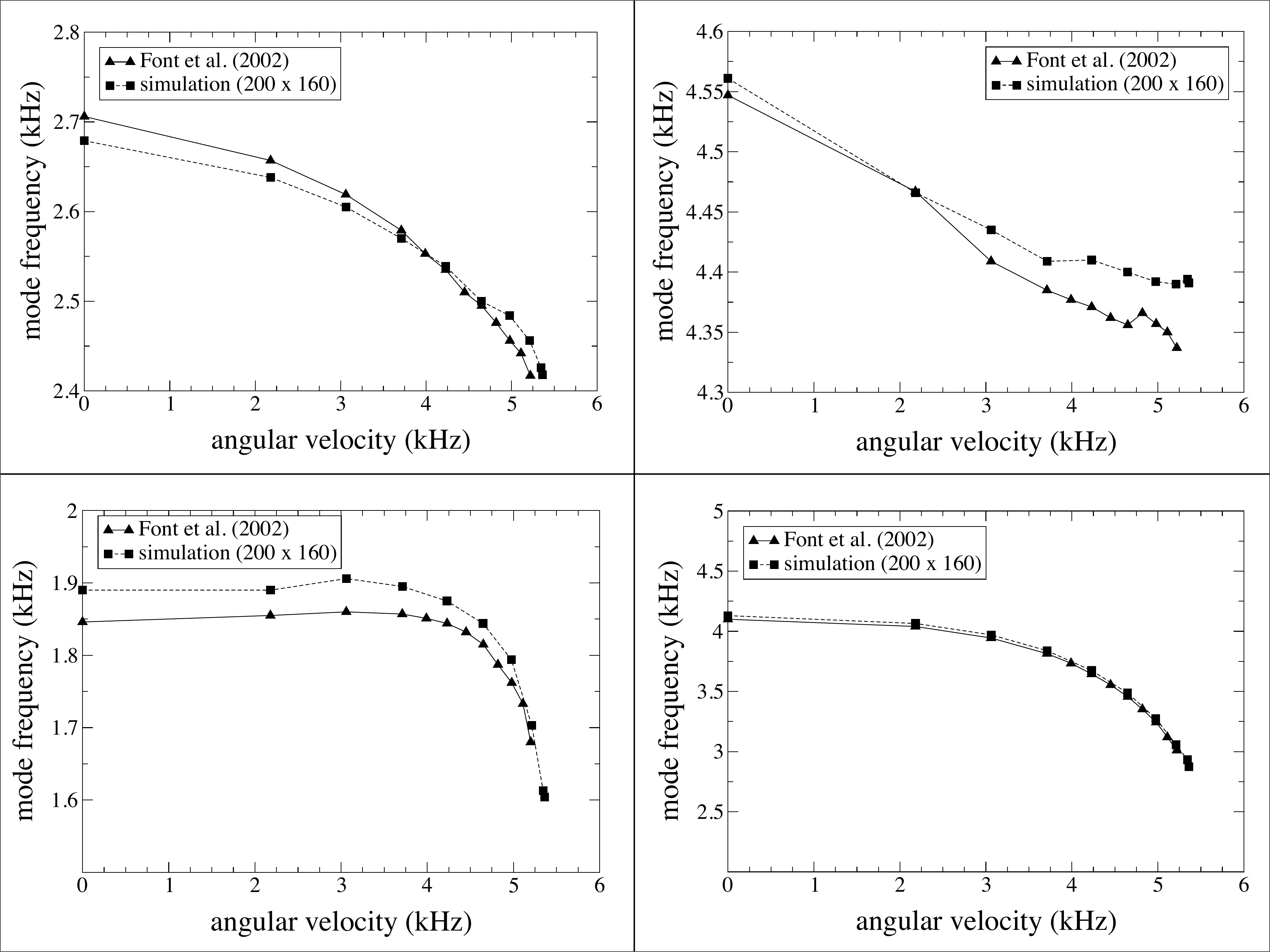} 
\caption{{\it Top Row:} The change in frequencies of the $F$- and $H_{1}\,$-mode for three different resolutions {\it Bottom Row:} The corresponding changes of the $^{2}f$- and $^{2}p_{1}\,$-mode}
\label{fig:resolutionDependence}
\end{figure}

%%%%%%%%%%%%%%%%%%%%%%%%%%%%%%
\subsubsection{Axial perturbations}
\label{sssec:axiAndAxial}

The axial perturbations of the fluid correspond to the inertial modes, in this case small axial deviations from equilibrium are restored by the  Coriolis force. These modes degenerate at zero frequency but have nonzero frequency once rotation sets in.
We were  able to identify some  inertial modes and to compare our results with the corrsponding studies in \cite{Dimmelmeier:2005zk}. In contrast to the Cowling-approximation we use here, they developed a full general relativistic hydrodynamics code under the assumption of a conformally flat three-metric and tested it on non-linear axisymmetric pulsations of rotating relativistic stars. Nevertheless, as we show, the results of the two calculations agree quite well.

\begin{figure}[htp!]
\centering
\includegraphics[width=3in]{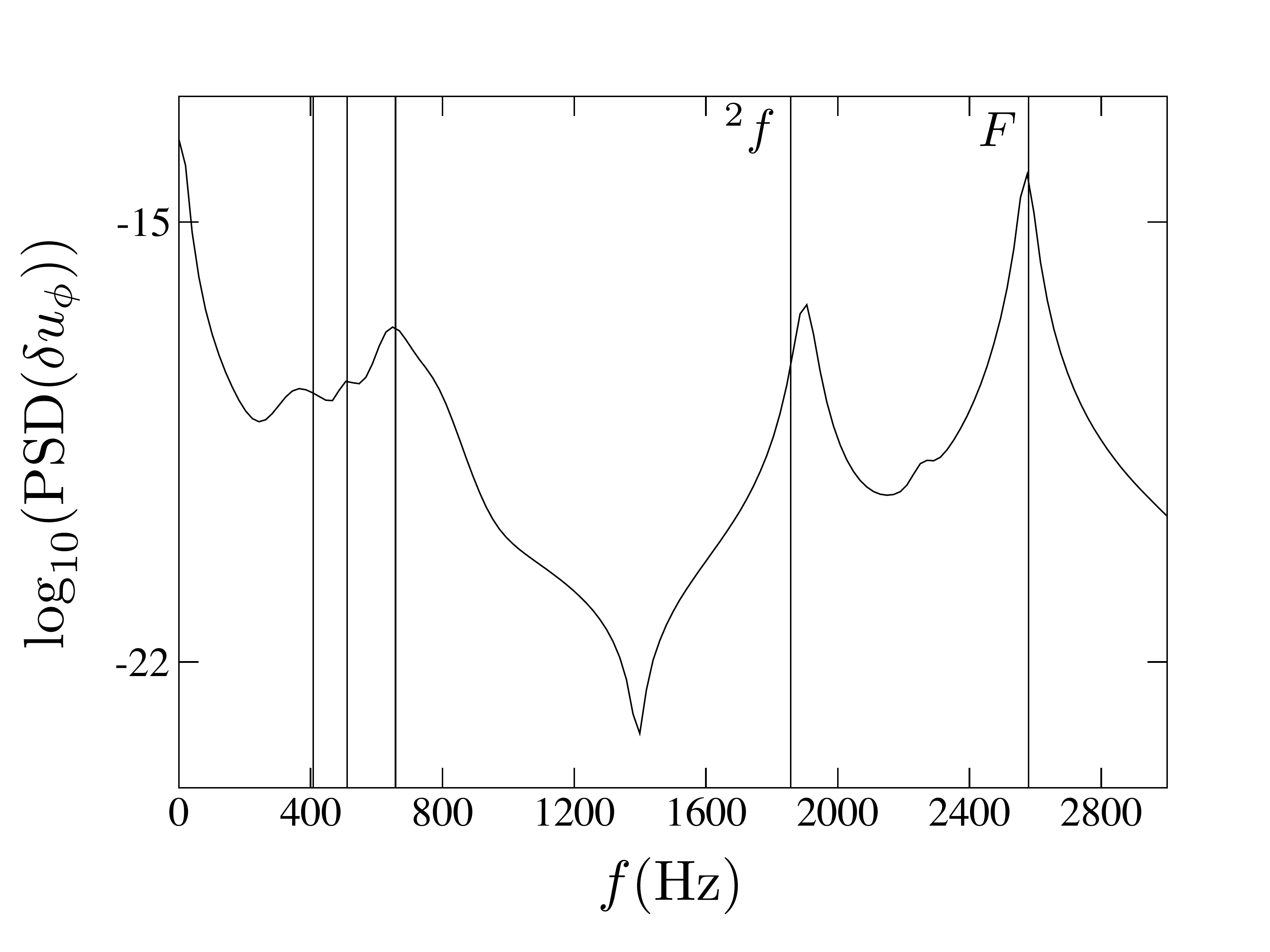} 
\caption{Fourier transform of $u_{\varphi}$ from a time series with BU3 as background model. The three vertical lines are from left to right the $i_{-2}$, $i_{1}$ and $i_{2}$ inertial mode frequencies listed in \cite{Dimmelmeier:2005zk}}
\label{fig:rModesBU}
\end{figure}

Figure \ref{fig:rModesBU} shows the power spectral density of the $u_{\varphi}$ velocity-component of a simulation with a BU3 background model (in the notation of \cite{Dimmelmeier:2005zk}). This neutron star has a ratio of polar coordinate radius to equatorial coordinate radius of $r_{p}/r_{e} = 0.85$ and rotates with an angular velocity $\Omega = 3.71$ kHz. We chose to monitor the $\varphi$-component of the velocity since this is the quantity where the inertial mode signature is typically well pronounced. The three vertical lines in the low frequency part of the plot denote the values of the $i_{-2}$, $i_{1}$ and $i_{2}$ mode frequency according to \cite{Dimmelmeier:2005zk}; on the right side one can identify the frequency peaks of the $^{2}f$- and the $F$-mode from table \ref{tb:BUmodelAxisymmetric1} and \ref{tb:BUmodelAxisymmetric2}.

\noindent
One can see, that there is a quite good agreement between our results and the literature values for this specific model and rotation rate; there is a greater deviation from the results in \cite{Dimmelmeier:2005zk} for the lowest frequency inertial mode $i_{-2}$. In general, the $i_{2}$ mode is the strongest inertial mode in our simulation and if one compares the various r-mode frequencies for different angular velocities, we find the best matching results for the $i_{1}$ and $i_{2}$ mode and still a very good match for the $i_{-2}$ frequencies. This is depicted in the next figure \ref{fig:rModesBUcompare}.

\begin{figure}[htp!] 
\centering
\includegraphics[width=0.34\textwidth]{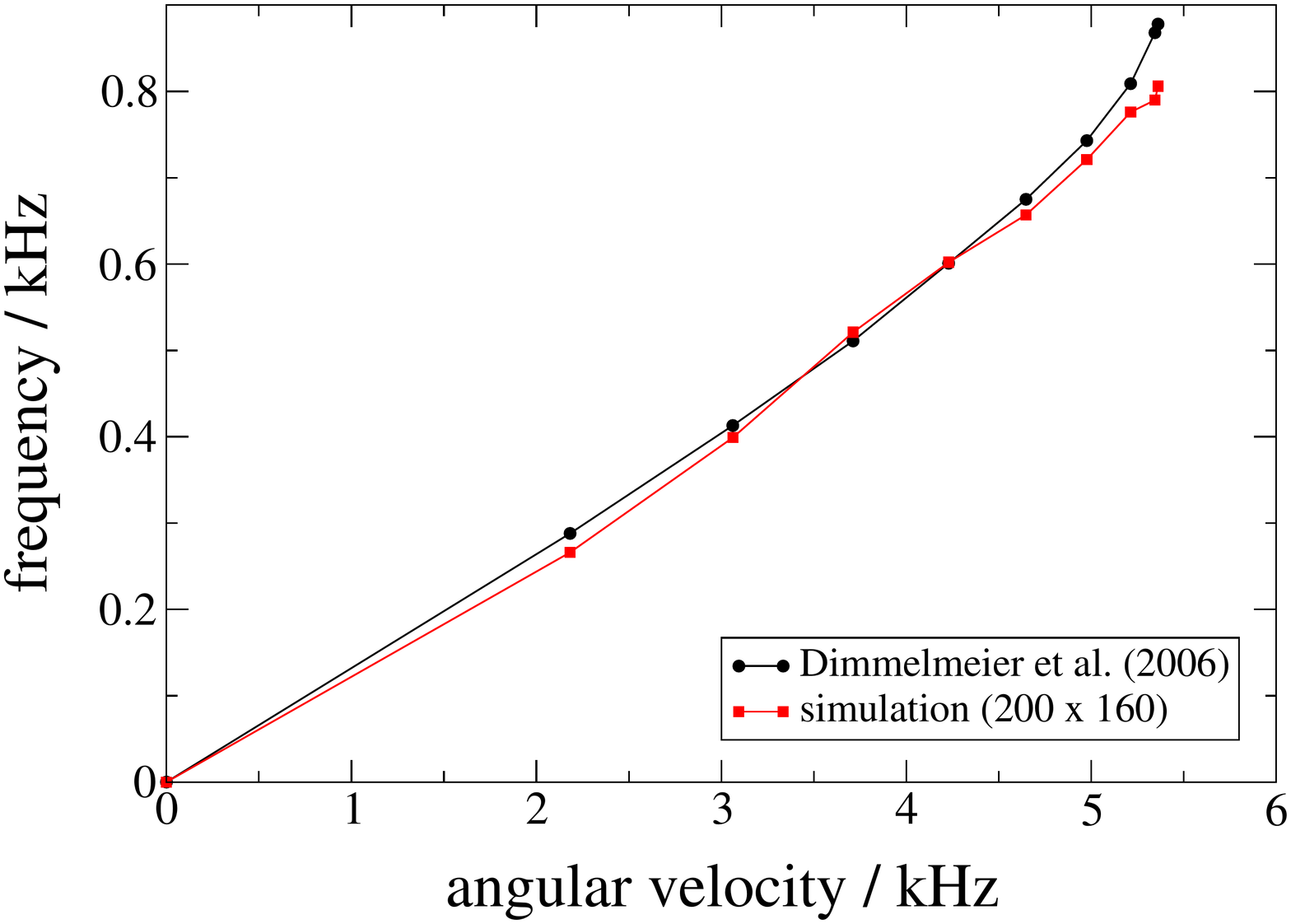}
\hspace{-0.2in}
\includegraphics[width=0.34\textwidth]{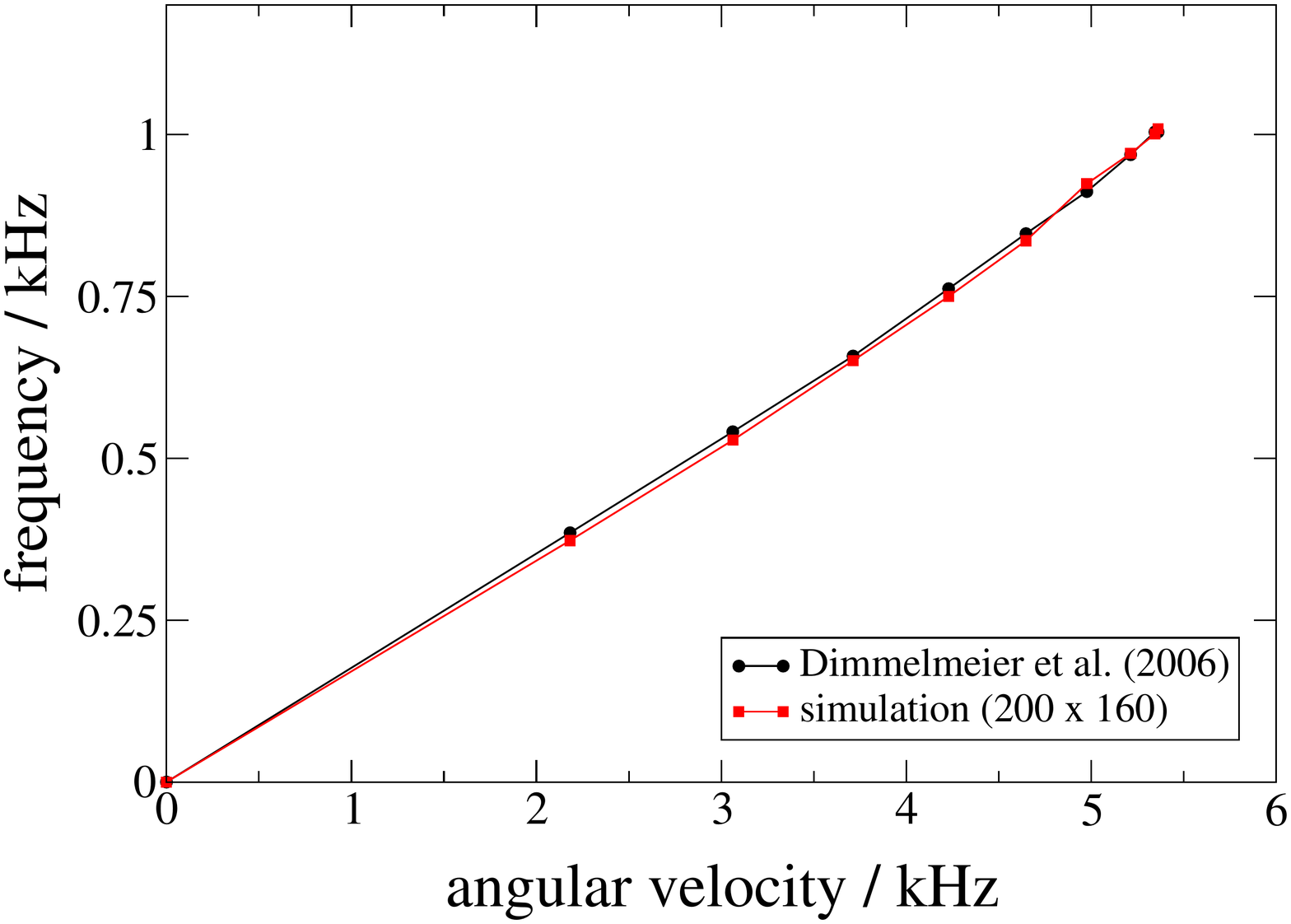}
\hspace{-0.2in}
\includegraphics[width=0.34\textwidth]{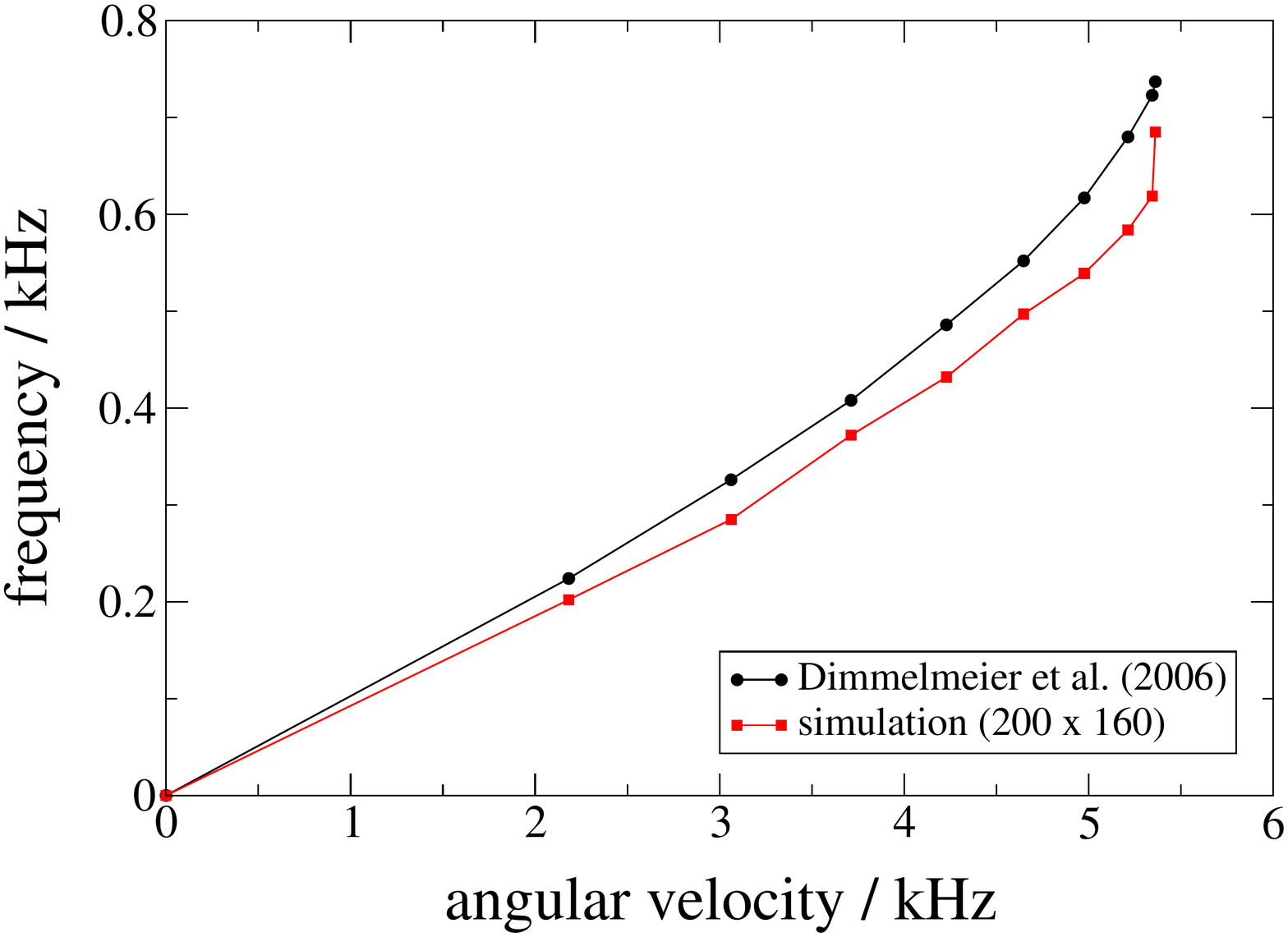}
\caption{A comparison between the  values in \cite{Dimmelmeier:2005zk} and our results for the $i_{1}$, $i_{2}$ and $i_{-2}$ inertial mode. Our simulations were performed with $200\times 160$ gridpoints}
\label{fig:rModesBUcompare}
\end{figure}
\noindent
So although we are using the Cowling approximation and not a full relativistic treatment, we can confirm the dependence of the three inertial modes  $i_{1}$, $i_{2}$ and $i_{-2}$ on the rotation rate described in the literature; a summary of our results is given in table \ref{tb:rModesBU}.

\begin{table}[ht!]
\begin{tabular}{| c | c | c  | c | c |}
\hline
~~Model~~ & ~~$\Omega$ (kHz)~~ & ~~$i_{1}$ (kHz)~~ & ~~$i_{2}$ (kHz)~~ & ~~ $i_{-2}$ (kHz)\\
\hline
BU0		& 0.0     	& 0.0	& 0.0	& 0.0\\
BU1	 	& 2.182   	& 0.266	& 0.373	& 0.202\\
BU2		& 3.062   	& 0.399	& 0.528	& 0.285\\
BU3		& 3.712   	& 0.521	& 0.651	& 0.372\\
BU4		& 4.229   	& 0.602	& 0.750	& 0.432\\
BU5		& 4.647   	& 0.657	& 0.836	& 0.497\\
BU6		& 4.976   	& 0.721	& 0.924	& 0.539\\
BU7		& 5.213   	& 0.776	& 0.971	& 0.584\\
BU8		& 5.344   	& 0.790	& 1.001	& 0.619\\
BU9		& 5.361	& 0.806	& 1.009	& 0.685\\
\hline
\end{tabular}
\caption{Frequencies of the three inertial modes $i_{1}$, $i_{2}$, $i_{-2}$ for the BU model at different rotation rates}
\label{tb:rModesBU}
\end{table}
%%%%%%%%%%%%%%

\subsection{Non-axisymmetric case}
\begin{figure}[htp!]
\centering
\includegraphics[width=3.5in]{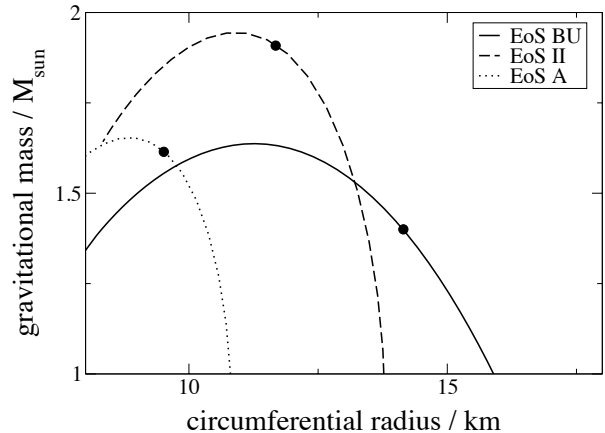} 
\caption{Mass-Radius relations for the EoS used to study non-axisymmetric perturbations; the black dots denote our actual models}
\label{fig:massRadiusDiagram}
\end{figure}

%%%%%%%%%
We will now turn to the study of non-axisymmetric oscillations on rotating compact objects with emphasis on the $m=2$-perturbations. In addition to the equation of state for the BU model series we used in the previous section mostly for code testing purposes, here we will apply two more equations of state. We use the polytropic parameters for EOS A and EOS II from \cite{2004PhRvD..70h4026S}; more specifically we have $\Gamma = 2.46$ and $K = 0.00936$ for EOS A and $\Gamma = 2.34$ and $K = 0.0195$ for EOS II. These values are given in geometric units ($G = M_{\odot} = 1$) and with [km] as length scale. Figure \ref{fig:massRadiusDiagram} shows mass-radius--diagrams for all the EoS we are using in this paper.
\noindent
The black dots on every of these three curves are the actual models used in the simulations. The BU model is a ``standard" compact object with a mass of $M = 1.4\,M_{\odot}$ and a circumferential radius of $R = 14.15$ km as one can see from the figure. In contrast to this the two other configurations we chose are very close to their maximum mass limit; in particular our background model for EOS A has a mass of $M = 1.61\,M_{\odot}$ and a radius of $R = 9.51$ km while the EOS II model has a mass of $M = 1.91\,M_{\odot}$ and a radius of $R = 11.68$ km. They are therefore more compact and their Kepler-limit is much higher than for the BU model; we will see what this means for the non-axisymmetric mode frequencies in the following discussion.
\subsubsection{Polar perturbations}
\label{sssec:nonaxiAndPolar}
\begin{figure}[htp!]
\centering
\includegraphics[width=3.5in]{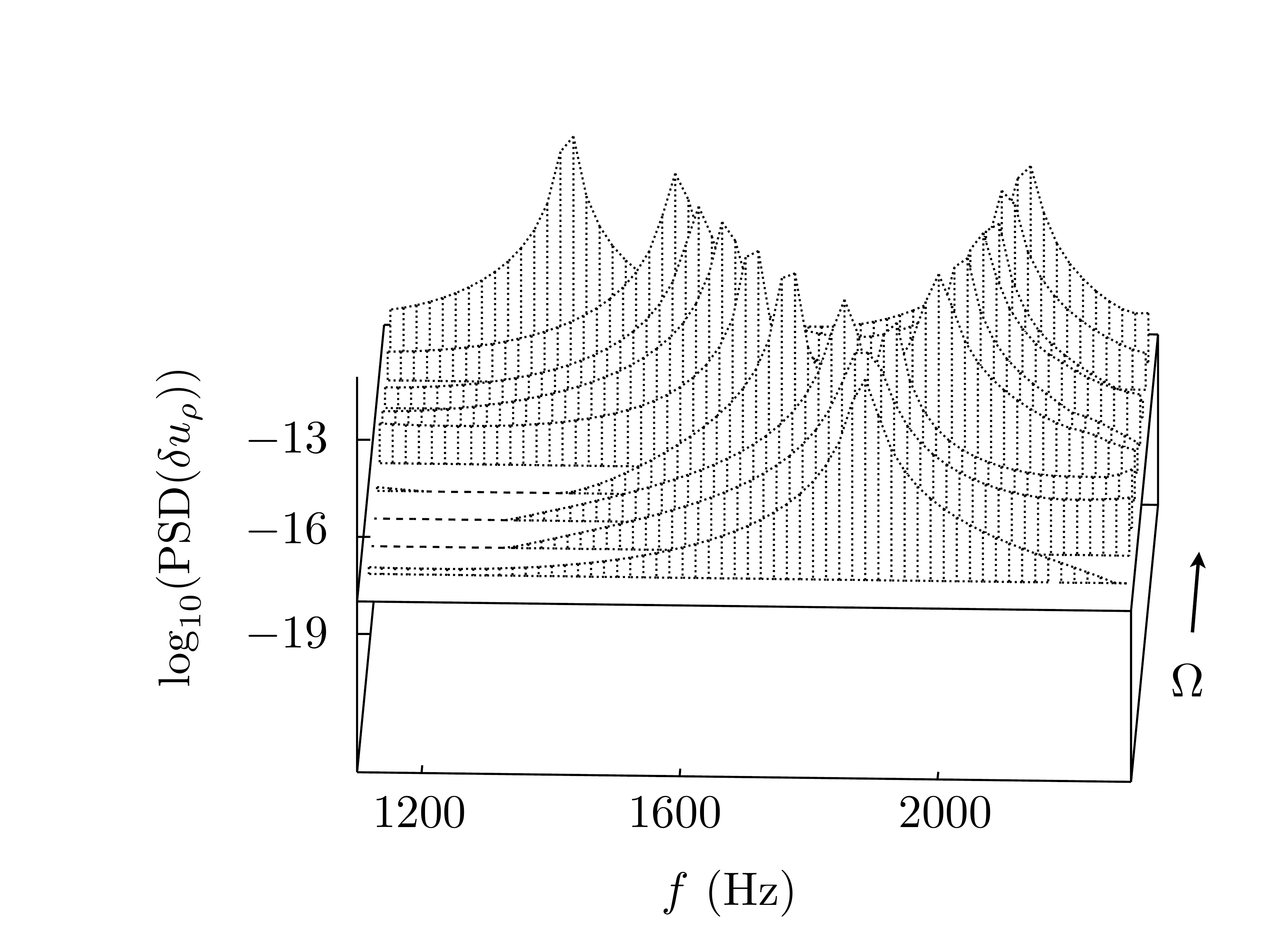} 
\caption{Power Spectral density of the $m=2$ $f$-mode for initial data provided by \eqref{eq:velocityPerturbation}. The splitting in the spectrum becomes apparent for increasing angular velocities}
\label{fig:mEquals2_snapshots}
\end{figure}

 The procedure for the excitation of modes is similar to the one in the axisymmetric case. Similar to the approach taken in \cite{Font:2001eu} we chose a $l=2$-velocity perturbation of the form
\begin{equation}
\label{eq:velocityPerturbation}
\delta u_{\theta} = A\sin\left(\frac{\pi r}{r_{s}(\theta)}\right)\sin\theta\cos\theta
\end{equation}
with the same meaning of $A$ and $r_{s}(\theta)$ as in \eqref{eq:lEqualsZeroPert}. Since we are working in cylindrical coordinates we have to decompose this perturbation into its $\rho$- and $\zeta$-component before we can insert it in our simulation. Figure \ref{fig:mEquals2_snapshots} shows a series of power spectral density plots for the BU model at different rotation rates.
\noindent
The frontmost spectrum shows the f-mode peak for a nonrotating star while the last one has been extracted for a star with a ratio of polar coordinate radius to equatorial coordinate radius of $r_{p}/r_{e} = 0.9$. In the nonrotating case the $f$-mode frequency is degenerated, i.e. it shows only one peak for the $m=2$ (counter-rotating) and the $m=-2$ (co-rotating) modes. This degeneracy is broken as soon as rotation sets in; for $\Omega \neq 0$ this peak splits into two peaks and breaks the symmetry between counter- and co-rotating modes. The very same behaviour can be observed for other modes as well.

\begin{figure}[htp!] 
\centering
\includegraphics[width=0.5\textwidth]{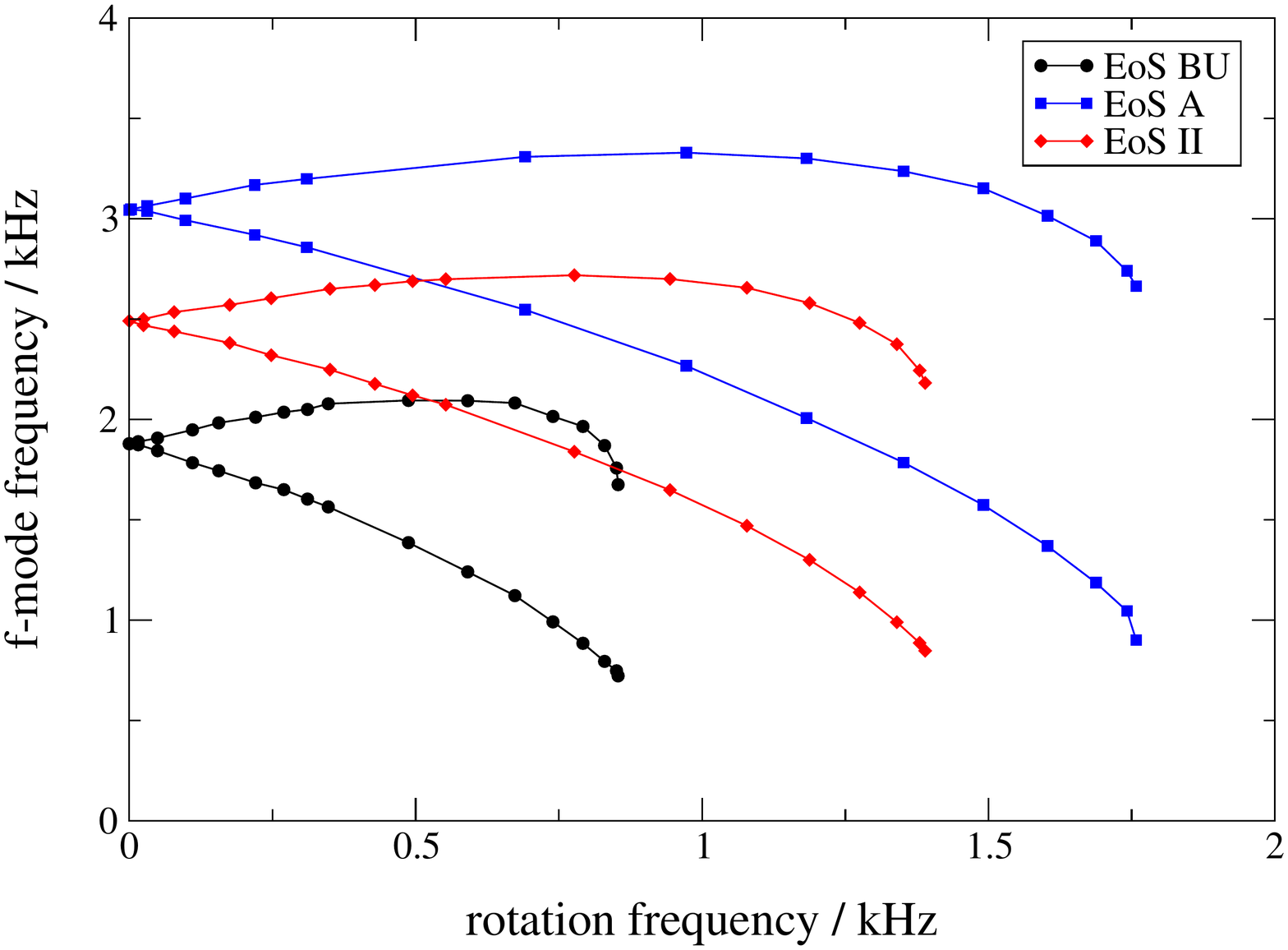}
\hspace{-0.2in}
\includegraphics[width=0.5\textwidth]{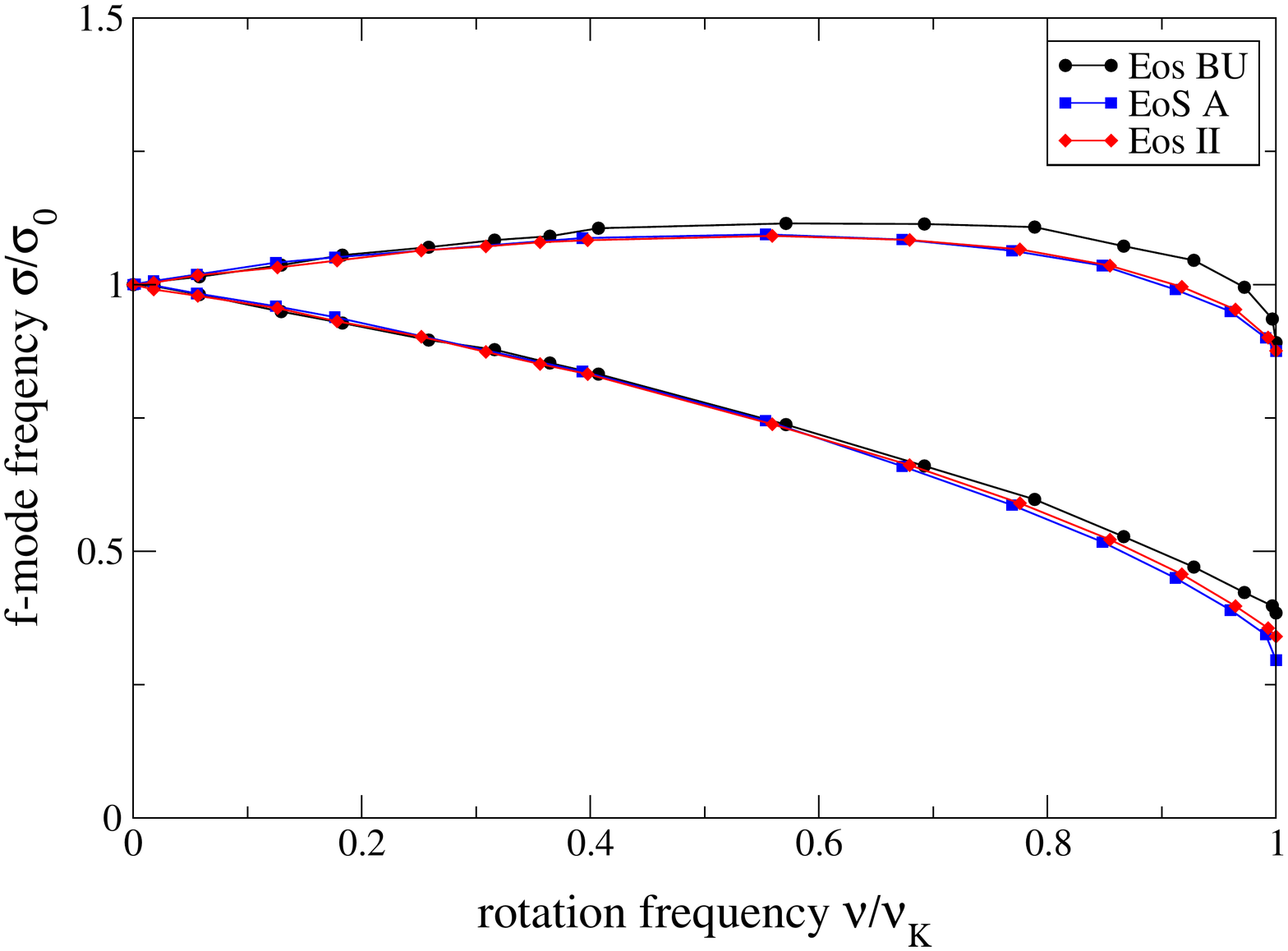}
\hspace{-0.2in}
\caption{The f-mode frequencies for the three models in consideration; in the left panel they are normalized with the corresponding Kepler rotation frequency $\nu_K$.}
\label{fig:fModeComov}
\end{figure}

In the following discussion we will focus on the fundamental mode although this particular mode suffers most of all from the Cowling approximation (compare \cite{Font:2001eu} with \cite{Dimmelmeier:2005zk} or see \cite{1968AnAp...31..549R} for an early Newtonian calculation). However in this paper we are interested in the evolution of the $f$-mode frequency with increasing angular velocity. While the absolute values may be incorrect by a factor of 20-30\% we still should be able to make some statements about the characteristic behaviour of the $f$-mode frequency dependence on the equations of state. Figure \ref{fig:fModeComov} shows our results for the models depicted in figure \ref{fig:massRadiusDiagram}. In contrast to previous figures we now plot the oscillation frequencies against the rotation frequency which is given by $\nu = \Omega/2\pi$.
%%%%%%%%%%
\noindent
As one can see the various models have a quite different range of f-mode frequencies. The BU model which is the less compact also has the lowest fundamental frequency. Due to the model parameters the Kepler limit for this particular neutron star is reached already at $\nu_{K} = 853$ Hz. The models for the other equations of state are more compact and therefore allow for higher rotation rates which can be as high as $\nu = 1.758$ kHz for EOS A. For each of the three EoS the $m=2$ branches are those in figure \ref{fig:fModeComov} with the higher frequencies and we will see in a second how we have been able to determine this. The frequency of the fundamental mode scales with the compactness of the star, higher compactness means higher frequency; a property we can directly validate from the left panel of figure \ref{fig:fModeComov}.

The right panel shows a different representation of the same picture where we plot for each model the normalized mode frequency $\sigma/\sigma_{0}$ against the normalized rotation frequeny $\nu/\nu_{K}$ where $\sigma_{0}$ is the mode frequency in the nonrotating limit and $\nu_{K}$ labels the Kepler limit for the rotation frequency. It is quite remarkable that although the models used in these simulations have very different parameters their normalized $f$-mode frequencies change nearly in the same manner when rotation is increased. It is only in the regime close to the Kepler limit where deviations for the various models become evident. For all rotation parameters we can write
\begin{equation}
\label{eq:linearApprox}
\frac{\sigma}{\sigma_{0}} = 1.0 + C_{lm}^{(1)}\left(\frac{\nu}{\nu_{K}}\right) + C_{lm}^{(2)}\left(\frac{\nu}{\nu_{K}}\right)^2
\end{equation}
independent of the specific EoS. To determine the coefficients $C_{lm}$ we made least-square fittings of the all data points we obtained from the various simulations. In particular we find $C_{22}^{(1)} = -0.25\pm 0.02$ and $C_{22}^{(2)} = -0.38 \pm 0.02$ for the $m=2$ solution and $C_{2 -2}^{(1)} = 0.48\pm 0.03$ and $C_{2-2}^{(2)} = -0.55 \pm 0.04$ in the $m=-2$ case. 

However, one should keep in mind that the results presented up to now are all derived in the corotating frame since this is the natural coordinate system in which our equations were formulated (see section \ref{sec:numMethod}). The only coordinate that changes when going to a stationary coordinate system is the azimuthal angle $\varphi$ and the relation connecting these two coordinates simply is
\begin{equation}
\label{eq:phiTransform}
\varphi_{\mathrm{corot}} = \varphi_{\mathrm{stat}} - \Omega t
\end{equation}
Due to the decomposition \eqref{eq:perturbedAnsatz} of our perturbation variables and the harmonic Fourier transformation we are performing on our numerically obtained time-series, we are effectively decomposing our time-evolution quantities like 
\begin{equation}
\label{eq:completeDecomposition}
f\sim e^{i\sigma t}e^{im\varphi} = e^{i(\sigma t + m\varphi)}
\end{equation}
To track a specific constant phase in time one therefore has to move on by an angle
\begin{equation*}
\varphi_{0}^{\mathrm{corot}} = -\frac{\sigma}{m}t_{0} 
\end{equation*}
after a time $t_{0}$. This means that in this case modes with a positive $m$ are moving retrograde while waves with a negative $m$ travel prograde in the comoving frame. We now insert \eqref{eq:phiTransform} into \eqref{eq:completeDecomposition} to obtain a relationship between the mode frequencies in the comoving and stationary frame and arrive at
\begin{equation}
\label{eq:frequencyRelation}
\sigma_{\mathrm{stat}} = \sigma_{\mathrm{corot}} - m\Omega
\end{equation}
For axisymmetric perturbations (i.e. $m=0$) the two frequencies are identical; this is why we didn't start this discussion already in section \ref{ssec:Axi}. To track a surface of constant phase in the stationary frame a similar calculation like the one above yields
\begin{equation*}
\varphi_{0}^{\mathrm{stat}} = -\frac{(\sigma_{\mathrm{corot}} - m \Omega)}{m}t_{0}
\end{equation*}
This means that if the frequency $\sigma_{\mathrm{corot}}$ is larger than $m\Omega$, then the mode is also travelling in retrograde direction in the stationary system. For $\sigma_{\mathrm{corot}} = m\Omega$ the frequency becomes degenerate in this system while for $\sigma_{\mathrm{corot}} < m\Omega$ a mode travelling retrograde in the comoving frame is seen prograde in the stationary coordinate system. The next figure shows our results for the stationary frame.

\begin{figure}[htp!] 
\centering
\includegraphics[width=0.5\textwidth]{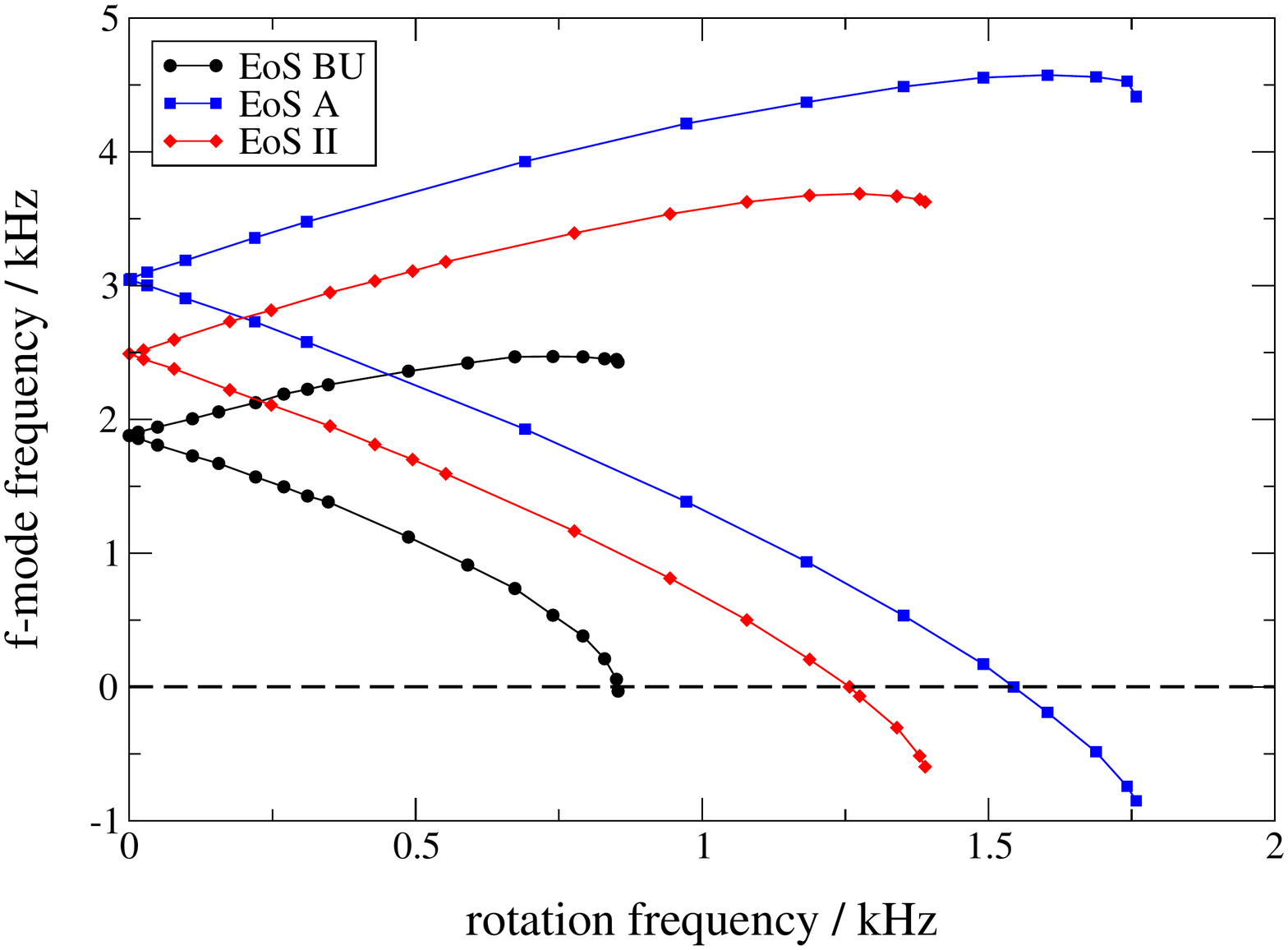}
\hspace{-0.2in}
\includegraphics[width=0.5\textwidth]{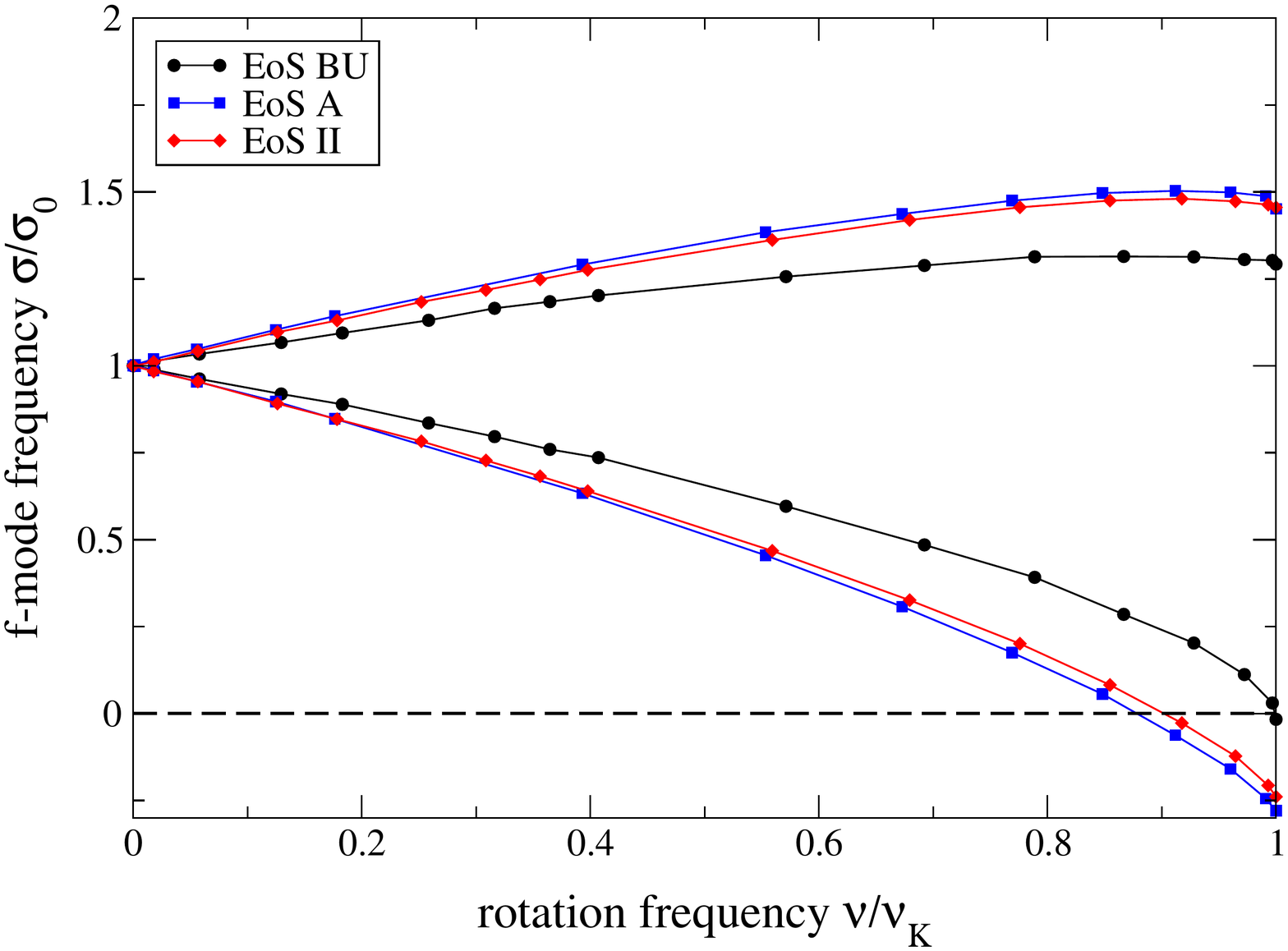}
\hspace{-0.2in}
\caption{Same as in figure \ref{fig:fModeComov} but now in the stationary frame }
\label{fig:fModeStat}
\end{figure}
\noindent
When we analyze the various fourier spectra in the stationary frame, we can actually see that for every of the three EoS the high frequency branches in figure \ref{fig:fModeComov} are shifted towards lower frequencies; together with equation \eqref{eq:frequencyRelation} it means that this branches can be identified with the $m=2$ solutions and vice versa. This is how we can identify the different sections of the curves.
The $m=2$ solutions of all models actually reach the point where $\sigma$ is zero; for the model BU this happens just at the Kepler limit, for the other models which are more compact this occurs even earlier. Beyond this point the $f$-mode is seen retrograde in the comoving frame but prograde in the stationary coordinate system. In this case the mode becomes CFS unstable and it can be an excellent source for gravitational waves. Finally, the normalized picture on the right panel now differs significantly from the corresponding picture in the corotating frame. This is due to the fact that the transformation from one system to the other introduces extra terms which are obviously model-dependent.

\subsubsection{Axial perturbations}
%%%%%%%%%%%%%%%%%%%%%%%%%
Non-axisymmetric axial perturbations, also known as r-modes, are known to be generically unstable to the CFS-instability at all rotation rates. This is an exciting class of stellar oscillations with many possible applications in astrophysics and gravitational wave research \cite{2001IJMPD..10..381A}.

We also did a couple of simulations to specifically excite the $l=2, m=2$ inertial mode and were successful. In the Newtonian framework, the fundamental r-mode is of purely axial parity and thus does not mix with high order polar terms, see \cite{1999ApJ...521..764L}. It also has been shown in \cite{Lockitch:2003jt}, that the contribution of higher order terms introduced by the fully general relativistic treatment can be neglected in the case of slow rotation. As for the polar non-axisymmetric oscillations we start with a perturbation of the $\theta$-component of the 4-velocity and expect a dependence $\sim 1/r\sin(\theta)$ of $\delta u_{\theta}$, see also \cite{Boutloukos:2006dz}. Typically, several mode recycling runs are needed to get a sharp and clean signal in the power spectral density. Also, due to the low frequencies of inertial modes especially at small rotation rates one needs much longer evolution times than for pressure driven modes. We chose to cancel the time-evolution after roughly $50$ ms to extract the eigenfrequencies. The numerical code is still stable there and can in principle evolve the initial data for a longer time intervall, leading to a more accurate frequency determination. However, we found that an evolution time of about $50$ ms and a spatial resolution of $200\times 160$ gridpoints is already quite good for a first estimation; there is only a marginally change when using longer integration intervals. We compare our code with results for a BU6 star rotating at $93\%$ of its maximum speed as described in \cite{Stergioulas:2000vs} where a nonlinear general-relativistic code has been used to study the saturation amplitude of r-modes. The following figure \ref{fig:nonaxialrmode} gives a summary of the results. 

\begin{figure}[htp!]
\centering
\includegraphics[width=5.2in]{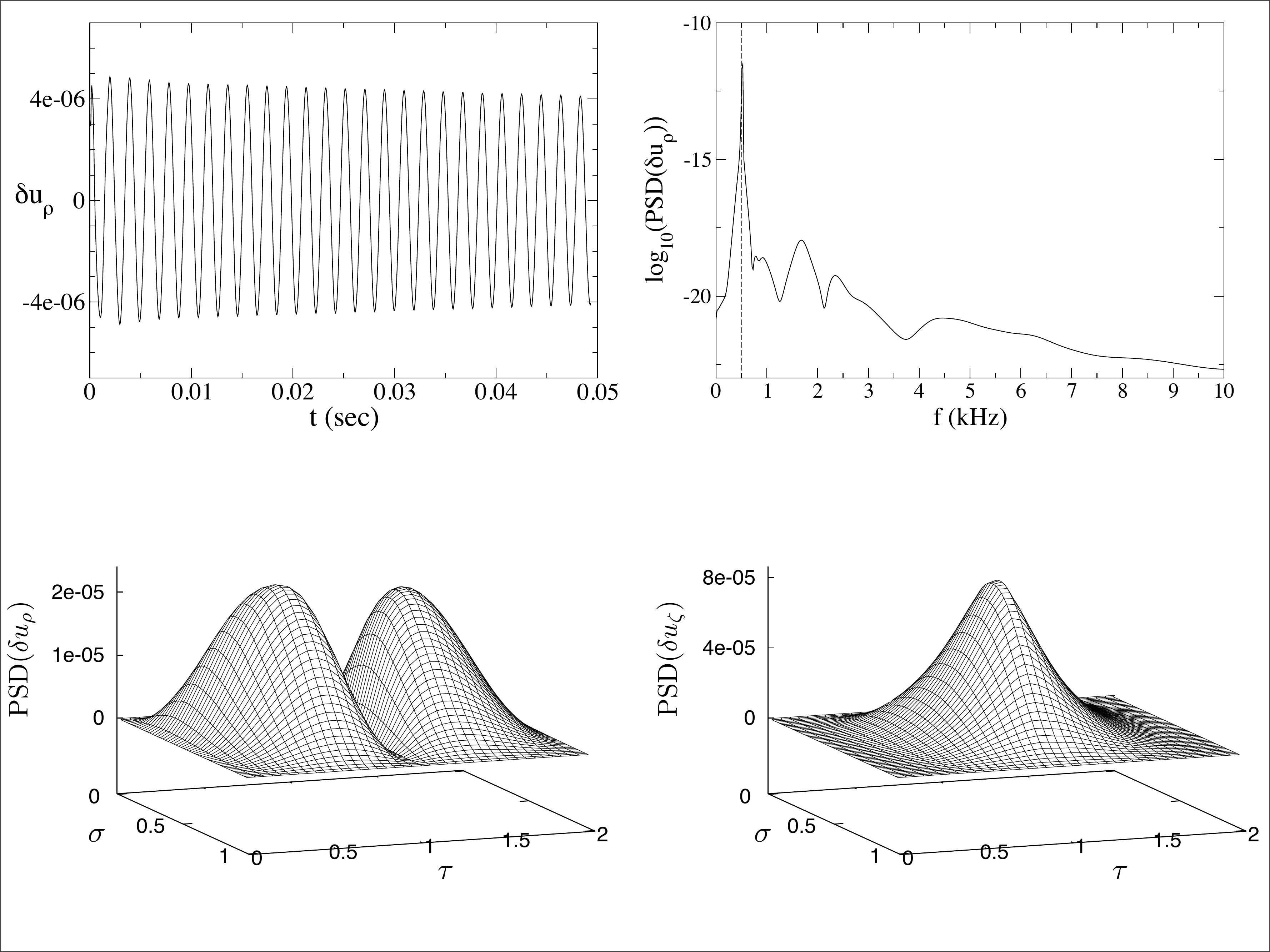} 
\caption{{\it Top Left:} Time evolution of $\delta u_{\rho}$ {\it Top Right:} Corresponding power spectral density plot {\it Bottom Left:} Shape of the $\delta u_{\rho}$ eigenfunction {\it Bottom Right:} Shape of the $\delta u_{\zeta}$ eigenfunction}
\label{fig:nonaxialrmode}
\end{figure}
%\noindent
By means of repeated mode recycling, the fundamental r-mode can be significantly enhanced as depicted in the PSD plot of figure \ref{fig:nonaxialrmode}. We also checked the angular dependence of the extracted eigenfunction. Since $\delta u_{\theta}$ should be $\sim\sin(\theta)$, the $\rho$- and $\zeta$-components of the perturbed 4-velocity are modified by an additional factor of $\cos(\theta)$ and $\sin(\theta)$ respectively. Keep in mind that the grid variable $\tau$ which takes values from $1$ to $2$ can be thought of similar to the polar angle $\theta$ in spherical coordinates; see the discussion of figure \ref{fig:akmGrid} in this paper. Then it is clear that the eigenfunction for $\delta u_{\rho}$ and $\delta u_{\zeta}$ depicted in the bottom row of figure \ref{fig:nonaxialrmode} show indeed the expected behaviour. The fundamental r-mode has a frequency of $f_{c} = 518$ Hz in the  comoving frame which translates to a frequency of $f_{i} = 1.066$ kHz in the inertial frame. This is in excellent agreement with \cite{Stergioulas:2000vs} where they found it at $f_{i} = 1.03$ kHz and also with \cite{Yoshida:2004gk} where they saw the mode at $f_{i} = 1.05$ kHz; nonlinear effects in the simulations of \cite{Stergioulas:2000vs} may explain the larger discrepancy there.

%%%%%%%%%%%%%%%%%%%%%%%
\section{Conclusions}

In this work we present a study of the oscillation properties of fast rotating relativistic stars for bot both axisymmetric and non-axisymmetric perturbations. The study was based on the  Cowling approximation  using a 2D version of the perturbation equations. The results for axisymmetric perturbations are in excellent agreement with earlier ones while the non-axisymmetric results are the first of their kind in the literature. We demonstrated the neutral points for the onset of the CFS instability and suggested possible normalizations which can be used in order to extract the parameters of the rotating star.  

The method  presented here (evolution of the 2D linearized equations) can be extended by including the perturbations of the spacetime. This will offer the possibility in testing the earlier results \cite{Stergioulas:1997ja} for the onset of the secular instability in fast rotating stars. Moreover, the  dependence of oscillations frequencies on rotation will be based on the exact results and will not rely on the Cowling approximation.  Finally, the effect of differential rotation on the spectra can be studied both for testing earlier, mainly Newtonian, results \cite{PhysRevD.64.024003,  2002ApJ...568L..41Y} but more importantly in finding the correct dependence of the frequencies on rotation as well as the neutral points for the onset of the secular instabilities. This is actually an important step since newly born neutron stars are expected to rotate differentially at least during the stages that they will be in an oscillatory phase or the even when they are secularly unstable.

%%%%%%%%%%%%%%%%%%%%%%%%  ACKNOWLEDGEMENTS  %%%%%%%%%%%%%%%%%%%%%%%%%%%%%%%%%%%%
\[ \]
{\bf Acknowledgments:} We thank D. Petroff and N. Stergioulas  for helpful discussions.  
This work was also supported by the German Science Foundation (DFG) via SFB/TR7.

%%%%%%%%%%%%%%%%%%%%%%%%%%%%%%%%%  BIBLIOGRAPHY  %%%%%%%%%%%%%%%%%%%%%%%%%%%%%%%
%\nocite*
\bibliographystyle{apsrev} \bibliography{references}

\begin{thebibliography}{65}
\expandafter\ifx\csname natexlab\endcsname\relax\def\natexlab#1{#1}\fi
\expandafter\ifx\csname bibnamefont\endcsname\relax
  \def\bibnamefont#1{#1}\fi
\expandafter\ifx\csname bibfnamefont\endcsname\relax
  \def\bibfnamefont#1{#1}\fi
\expandafter\ifx\csname citenamefont\endcsname\relax
  \def\citenamefont#1{#1}\fi
\expandafter\ifx\csname url\endcsname\relax
  \def\url#1{\texttt{#1}}\fi
\expandafter\ifx\csname urlprefix\endcsname\relax\def\urlprefix{URL }\fi
\providecommand{\bibinfo}[2]{#2}
\providecommand{\eprint}[2][]{\url{#2}}

\bibitem[{\citenamefont{Ott et~al.}(2007)}]{Ott:2006eh}
\bibinfo{author}{\bibfnamefont{C.~D.} \bibnamefont{Ott}} \bibnamefont{et~al.},
  \bibinfo{journal}{Class. Quant. Grav.} \textbf{\bibinfo{volume}{24}},
  \bibinfo{pages}{S139} (\bibinfo{year}{2007}).

\bibitem[{\citenamefont{Kokkotas and Schafer}(1995)}]{Kokkotas:1995xe}
\bibinfo{author}{\bibfnamefont{K.~D.} \bibnamefont{Kokkotas}} \bibnamefont{and}
  \bibinfo{author}{\bibfnamefont{G.}~\bibnamefont{Schafer}},
  \bibinfo{journal}{Mon. Not. Roy. Astron. Soc.}
  \textbf{\bibinfo{volume}{275}}, \bibinfo{pages}{301} (\bibinfo{year}{1995}).

\bibitem[{\citenamefont{Flanagan and Racine}(2007)}]{Flanagan:2006sb}
\bibinfo{author}{\bibfnamefont{E.~E.} \bibnamefont{Flanagan}} \bibnamefont{and}
  \bibinfo{author}{\bibfnamefont{E.}~\bibnamefont{Racine}},
  \bibinfo{journal}{Phys. Rev.} \textbf{\bibinfo{volume}{D75}},
  \bibinfo{pages}{044001} (\bibinfo{year}{2007}).

\bibitem[{\citenamefont{{Marranghello}
  et~al.}(2002)\citenamefont{{Marranghello}, {Vasconcellos}, and {de Freitas
  Pacheco}}}]{2002PhRvD..66f4027M}
\bibinfo{author}{\bibfnamefont{G.~F.} \bibnamefont{{Marranghello}}},
  \bibinfo{author}{\bibfnamefont{C.~A.} \bibnamefont{{Vasconcellos}}},
  \bibnamefont{and} \bibinfo{author}{\bibfnamefont{J.~A.} \bibnamefont{{de
  Freitas Pacheco}}}, \bibinfo{journal}{\prd} \textbf{\bibinfo{volume}{66}},
  \bibinfo{pages}{064027} (\bibinfo{year}{2002}).

\bibitem[{\citenamefont{Ferrari et~al.}(2007)\citenamefont{Ferrari, Gualtieri,
  and Pons}}]{Ferrari:2007ur}
\bibinfo{author}{\bibfnamefont{V.}~\bibnamefont{Ferrari}},
  \bibinfo{author}{\bibfnamefont{L.}~\bibnamefont{Gualtieri}},
  \bibnamefont{and} \bibinfo{author}{\bibfnamefont{J.~A.} \bibnamefont{Pons}},
  \bibinfo{journal}{Class. Quant. Grav.} \textbf{\bibinfo{volume}{24}},
  \bibinfo{pages}{5093} (\bibinfo{year}{2007}).

\bibitem[{\citenamefont{{McDermott} et~al.}(1988)\citenamefont{{McDermott},
  {van Horn}, and {Hansen}}}]{1988ApJ...325..725M}
\bibinfo{author}{\bibfnamefont{P.~N.} \bibnamefont{{McDermott}}},
  \bibinfo{author}{\bibfnamefont{H.~M.} \bibnamefont{{van Horn}}},
  \bibnamefont{and} \bibinfo{author}{\bibfnamefont{C.~J.}
  \bibnamefont{{Hansen}}}, \bibinfo{journal}{\apj}
  \textbf{\bibinfo{volume}{325}}, \bibinfo{pages}{725} (\bibinfo{year}{1988}).

\bibitem[{\citenamefont{{Kokkotas} and {Schmidt}}(1999)}]{1999LRR.....2....2K}
\bibinfo{author}{\bibfnamefont{K.~D.} \bibnamefont{{Kokkotas}}}
  \bibnamefont{and} \bibinfo{author}{\bibfnamefont{B.~G.}
  \bibnamefont{{Schmidt}}}, \bibinfo{journal}{Living Reviews in Relativity}
  \textbf{\bibinfo{volume}{2}}, \bibinfo{pages}{2} (\bibinfo{year}{1999}).

\bibitem[{\citenamefont{{Kokkotas} and {Schutz}}(1992)}]{Kokkotas:1992ks}
\bibinfo{author}{\bibfnamefont{K.~D.} \bibnamefont{{Kokkotas}}}
  \bibnamefont{and} \bibinfo{author}{\bibfnamefont{B.~F.}
  \bibnamefont{{Schutz}}}, \bibinfo{journal}{\mnras}
  \textbf{\bibinfo{volume}{255}}, \bibinfo{pages}{119} (\bibinfo{year}{1992}).

\bibitem[{\citenamefont{{Schumaker} and {Thorne}}(1983)}]{1983MNRAS.203..457S}
\bibinfo{author}{\bibfnamefont{B.~L.} \bibnamefont{{Schumaker}}}
  \bibnamefont{and} \bibinfo{author}{\bibfnamefont{K.~S.}
  \bibnamefont{{Thorne}}}, \bibinfo{journal}{\mnras}
  \textbf{\bibinfo{volume}{203}}, \bibinfo{pages}{457} (\bibinfo{year}{1983}).

\bibitem[{\citenamefont{{McDermott} et~al.}(1985)\citenamefont{{McDermott},
  {van Horn}, {Hansen}, and {Buland}}}]{1985ApJ...297L..37M}
\bibinfo{author}{\bibfnamefont{P.~N.} \bibnamefont{{McDermott}}},
  \bibinfo{author}{\bibfnamefont{H.~M.} \bibnamefont{{van Horn}}},
  \bibinfo{author}{\bibfnamefont{C.~J.} \bibnamefont{{Hansen}}},
  \bibnamefont{and} \bibinfo{author}{\bibfnamefont{R.}~\bibnamefont{{Buland}}},
  \bibinfo{journal}{\apjl} \textbf{\bibinfo{volume}{297}}, \bibinfo{pages}{L37}
  (\bibinfo{year}{1985}).

\bibitem[{\citenamefont{{Vavoulidis} et~al.}(2008)\citenamefont{{Vavoulidis},
  {Kokkotas}, and {Stavridis}}}]{2008MNRAS.384.1711V}
\bibinfo{author}{\bibfnamefont{M.}~\bibnamefont{{Vavoulidis}}},
  \bibinfo{author}{\bibfnamefont{K.~D.} \bibnamefont{{Kokkotas}}},
  \bibnamefont{and}
  \bibinfo{author}{\bibfnamefont{A.}~\bibnamefont{{Stavridis}}},
  \bibinfo{journal}{\mnras} \textbf{\bibinfo{volume}{384}},
  \bibinfo{pages}{1711} (\bibinfo{year}{2008}).

\bibitem[{\citenamefont{{Carroll} et~al.}(1986)\citenamefont{{Carroll},
  {Zweibel}, {Hansen}, {McDermott}, {Savedoff}, {Thomas}, and {van
  Horn}}}]{1986ApJ...305..767C}
\bibinfo{author}{\bibfnamefont{B.~W.} \bibnamefont{{Carroll}}},
  \bibinfo{author}{\bibfnamefont{E.~G.} \bibnamefont{{Zweibel}}},
  \bibinfo{author}{\bibfnamefont{C.~J.} \bibnamefont{{Hansen}}},
  \bibinfo{author}{\bibfnamefont{P.~N.} \bibnamefont{{McDermott}}},
  \bibinfo{author}{\bibfnamefont{M.~P.} \bibnamefont{{Savedoff}}},
  \bibinfo{author}{\bibfnamefont{J.~H.} \bibnamefont{{Thomas}}},
  \bibnamefont{and} \bibinfo{author}{\bibfnamefont{H.~M.} \bibnamefont{{van
  Horn}}}, \bibinfo{journal}{\apj} \textbf{\bibinfo{volume}{305}},
  \bibinfo{pages}{767} (\bibinfo{year}{1986}).

\bibitem[{\citenamefont{{Andersson} and
  {Kokkotas}}(2001)}]{2001IJMPD..10..381A}
\bibinfo{author}{\bibfnamefont{N.}~\bibnamefont{{Andersson}}} \bibnamefont{and}
  \bibinfo{author}{\bibfnamefont{K.~D.} \bibnamefont{{Kokkotas}}},
  \bibinfo{journal}{International Journal of Modern Physics D}
  \textbf{\bibinfo{volume}{10}}, \bibinfo{pages}{381} (\bibinfo{year}{2001}).

\bibitem[{\citenamefont{Stergioulas}(2003)}]{lrr-2003-3}
\bibinfo{author}{\bibfnamefont{N.}~\bibnamefont{Stergioulas}},
  \bibinfo{journal}{Living Reviews in Relativity} \textbf{\bibinfo{volume}{6}}
  (\bibinfo{year}{2003}),
  \urlprefix\url{http://www.livingreviews.org/lrr-2003-3}.

\bibitem[{\citenamefont{{Thorne} and {Campolattaro}}(1967)}]{Thorne:1967th}
\bibinfo{author}{\bibfnamefont{K.~S.} \bibnamefont{{Thorne}}} \bibnamefont{and}
  \bibinfo{author}{\bibfnamefont{A.}~\bibnamefont{{Campolattaro}}},
  \bibinfo{journal}{\apj} \textbf{\bibinfo{volume}{149}}, \bibinfo{pages}{591}
  (\bibinfo{year}{1967}).

\bibitem[{\citenamefont{{Price} and {Thorne}}(1969)}]{Price:1969pt}
\bibinfo{author}{\bibfnamefont{R.}~\bibnamefont{{Price}}} \bibnamefont{and}
  \bibinfo{author}{\bibfnamefont{K.~S.} \bibnamefont{{Thorne}}},
  \bibinfo{journal}{\apj} \textbf{\bibinfo{volume}{155}}, \bibinfo{pages}{163}
  (\bibinfo{year}{1969}).

\bibitem[{\citenamefont{{Thorne}}(1969{\natexlab{a}})}]{Thorne:1969to}
\bibinfo{author}{\bibfnamefont{K.~S.} \bibnamefont{{Thorne}}},
  \bibinfo{journal}{\apj} \textbf{\bibinfo{volume}{158}}, \bibinfo{pages}{1}
  (\bibinfo{year}{1969}{\natexlab{a}}).

\bibitem[{\citenamefont{{Thorne}}(1969{\natexlab{b}})}]{Thorne:1969th}
\bibinfo{author}{\bibfnamefont{K.~S.} \bibnamefont{{Thorne}}},
  \bibinfo{journal}{\apj} \textbf{\bibinfo{volume}{158}}, \bibinfo{pages}{997}
  (\bibinfo{year}{1969}{\natexlab{b}}).

\bibitem[{\citenamefont{{Detweiler} and {Lindblom}}(1985)}]{Detweiler:1985dl}
\bibinfo{author}{\bibfnamefont{S.}~\bibnamefont{{Detweiler}}} \bibnamefont{and}
  \bibinfo{author}{\bibfnamefont{L.}~\bibnamefont{{Lindblom}}},
  \bibinfo{journal}{\apj} \textbf{\bibinfo{volume}{292}}, \bibinfo{pages}{12}
  (\bibinfo{year}{1985}).

\bibitem[{\citenamefont{{Andersson} and {Kokkotas}}(1996)}]{Andersson:1996ak}
\bibinfo{author}{\bibfnamefont{N.}~\bibnamefont{{Andersson}}} \bibnamefont{and}
  \bibinfo{author}{\bibfnamefont{K.~D.} \bibnamefont{{Kokkotas}}},
  \bibinfo{journal}{\prl} \textbf{\bibinfo{volume}{77}}, \bibinfo{pages}{4134}
  (\bibinfo{year}{1996}).

\bibitem[{\citenamefont{{Andersson} and {Kokkotas}}(1998)}]{Andersson:1998ak}
\bibinfo{author}{\bibfnamefont{N.}~\bibnamefont{{Andersson}}} \bibnamefont{and}
  \bibinfo{author}{\bibfnamefont{K.~D.} \bibnamefont{{Kokkotas}}},
  \bibinfo{journal}{\mnras} \textbf{\bibinfo{volume}{297}},
  \bibinfo{pages}{493} (\bibinfo{year}{1998}).

\bibitem[{\citenamefont{Benhar et~al.}(1999)\citenamefont{Benhar, Berti, and
  Ferrari}}]{Benhar:1998au}
\bibinfo{author}{\bibfnamefont{O.}~\bibnamefont{Benhar}},
  \bibinfo{author}{\bibfnamefont{E.}~\bibnamefont{Berti}}, \bibnamefont{and}
  \bibinfo{author}{\bibfnamefont{V.}~\bibnamefont{Ferrari}},
  \bibinfo{journal}{Mon. Not. Roy. Astron. Soc.}
  \textbf{\bibinfo{volume}{310}}, \bibinfo{pages}{797} (\bibinfo{year}{1999}).

\bibitem[{\citenamefont{{Kokkotas} et~al.}(2001)\citenamefont{{Kokkotas},
  {Apostolatos}, and {Andersson}}}]{2001MNRAS.320...307K}
\bibinfo{author}{\bibfnamefont{K.~D.} \bibnamefont{{Kokkotas}}},
  \bibinfo{author}{\bibfnamefont{T.~A.} \bibnamefont{{Apostolatos}}},
  \bibnamefont{and}
  \bibinfo{author}{\bibfnamefont{N.}~\bibnamefont{{Andersson}}},
  \bibinfo{journal}{\mnras} \textbf{\bibinfo{volume}{320}},
  \bibinfo{pages}{307} (\bibinfo{year}{2001}).

\bibitem[{\citenamefont{Benhar et~al.}(2004)\citenamefont{Benhar, Ferrari, and
  Gualtieri}}]{benhar-2004-70}
\bibinfo{author}{\bibfnamefont{O.}~\bibnamefont{Benhar}},
  \bibinfo{author}{\bibfnamefont{V.}~\bibnamefont{Ferrari}}, \bibnamefont{and}
  \bibinfo{author}{\bibfnamefont{L.}~\bibnamefont{Gualtieri}},
  \bibinfo{journal}{Phys. Rev.} \textbf{\bibinfo{volume}{D70}},
  \bibinfo{pages}{124015} (\bibinfo{year}{2004}).

\bibitem[{\citenamefont{{Sotani} et~al.}(2004)\citenamefont{{Sotani}, {Kohri},
  and {Harada}}}]{2004PhRvD..69h4008S}
\bibinfo{author}{\bibfnamefont{H.}~\bibnamefont{{Sotani}}},
  \bibinfo{author}{\bibfnamefont{K.}~\bibnamefont{{Kohri}}}, \bibnamefont{and}
  \bibinfo{author}{\bibfnamefont{T.}~\bibnamefont{{Harada}}},
  \bibinfo{journal}{\prd} \textbf{\bibinfo{volume}{69}},
  \bibinfo{pages}{084008} (\bibinfo{year}{2004}).

\bibitem[{\citenamefont{{Sotani} and {Kokkotas}}(2004)}]{2004PhRvD..70h4026S}
\bibinfo{author}{\bibfnamefont{H.}~\bibnamefont{{Sotani}}} \bibnamefont{and}
  \bibinfo{author}{\bibfnamefont{K.~D.} \bibnamefont{{Kokkotas}}},
  \bibinfo{journal}{\prd} \textbf{\bibinfo{volume}{70}},
  \bibinfo{pages}{084026} (\bibinfo{year}{2004}).

\bibitem[{\citenamefont{{Unno} et~al.}(1989)\citenamefont{{Unno}, {Osaki},
  {Ando}, {Saio}, and {Shibahashi}}}]{1989nos..book.....U}
\bibinfo{author}{\bibfnamefont{W.}~\bibnamefont{{Unno}}},
  \bibinfo{author}{\bibfnamefont{Y.}~\bibnamefont{{Osaki}}},
  \bibinfo{author}{\bibfnamefont{H.}~\bibnamefont{{Ando}}},
  \bibinfo{author}{\bibfnamefont{H.}~\bibnamefont{{Saio}}}, \bibnamefont{and}
  \bibinfo{author}{\bibfnamefont{H.}~\bibnamefont{{Shibahashi}}},
  \emph{\bibinfo{title}{{Nonradial oscillations of stars}}}
  (\bibinfo{publisher}{Tokyo: University of Tokyo Press, 1989, 2nd ed.},
  \bibinfo{year}{1989}).

\bibitem[{\citenamefont{{Robe}}(1968)}]{1968AnAp...31..549R}
\bibinfo{author}{\bibfnamefont{H.}~\bibnamefont{{Robe}}},
  \bibinfo{journal}{Annales d'Astrophysique} \textbf{\bibinfo{volume}{31}},
  \bibinfo{pages}{549} (\bibinfo{year}{1968}).

\bibitem[{\citenamefont{Font et~al.}(2000)\citenamefont{Font, Stergioulas, and
  Kokkotas}}]{Font:1999wh}
\bibinfo{author}{\bibfnamefont{J.~A.} \bibnamefont{Font}},
  \bibinfo{author}{\bibfnamefont{N.}~\bibnamefont{Stergioulas}},
  \bibnamefont{and} \bibinfo{author}{\bibfnamefont{K.~D.}
  \bibnamefont{Kokkotas}}, \bibinfo{journal}{Mon. Not. Roy. Astron. Soc.}
  \textbf{\bibinfo{volume}{313}}, \bibinfo{pages}{678} (\bibinfo{year}{2000}).

\bibitem[{\citenamefont{{Stergioulas} and {Font}}(2001)}]{2001PhRvL..86.1148S}
\bibinfo{author}{\bibfnamefont{N.}~\bibnamefont{{Stergioulas}}}
  \bibnamefont{and} \bibinfo{author}{\bibfnamefont{J.~A.}
  \bibnamefont{{Font}}}, \bibinfo{journal}{Physical Review Letters}
  \textbf{\bibinfo{volume}{86}}, \bibinfo{pages}{1148} (\bibinfo{year}{2001}).

\bibitem[{\citenamefont{{Font} et~al.}(2001)\citenamefont{{Font},
  {Dimmelmeier}, {Gupta}, and {Stergioulas}}}]{Font:2001eu}
\bibinfo{author}{\bibfnamefont{J.~A.} \bibnamefont{{Font}}},
  \bibinfo{author}{\bibfnamefont{H.}~\bibnamefont{{Dimmelmeier}}},
  \bibinfo{author}{\bibfnamefont{A.}~\bibnamefont{{Gupta}}}, \bibnamefont{and}
  \bibinfo{author}{\bibfnamefont{N.}~\bibnamefont{{Stergioulas}}},
  \bibinfo{journal}{\mnras} \textbf{\bibinfo{volume}{325}},
  \bibinfo{pages}{1463} (\bibinfo{year}{2001}).

\bibitem[{\citenamefont{Stergioulas et~al.}(2004)\citenamefont{Stergioulas,
  Apostolatos, and Font}}]{Stergioulas:2003ep}
\bibinfo{author}{\bibfnamefont{N.}~\bibnamefont{Stergioulas}},
  \bibinfo{author}{\bibfnamefont{T.~A.} \bibnamefont{Apostolatos}},
  \bibnamefont{and} \bibinfo{author}{\bibfnamefont{J.~A.} \bibnamefont{Font}},
  \bibinfo{journal}{\mnras} \textbf{\bibinfo{volume}{352}},
  \bibinfo{pages}{1089} (\bibinfo{year}{2004}).

\bibitem[{\citenamefont{Dimmelmeier et~al.}(2006)\citenamefont{Dimmelmeier,
  Stergioulas, and Font}}]{Dimmelmeier:2005zk}
\bibinfo{author}{\bibfnamefont{H.}~\bibnamefont{Dimmelmeier}},
  \bibinfo{author}{\bibfnamefont{N.}~\bibnamefont{Stergioulas}},
  \bibnamefont{and} \bibinfo{author}{\bibfnamefont{J.~A.} \bibnamefont{Font}},
  \bibinfo{journal}{\mnras} \textbf{\bibinfo{volume}{368}},
  \bibinfo{pages}{1609} (\bibinfo{year}{2006}).

\bibitem[{\citenamefont{{Kastaun}}(2006)}]{2006PhRvD..74l4024K}
\bibinfo{author}{\bibfnamefont{W.}~\bibnamefont{{Kastaun}}},
  \bibinfo{journal}{\prd} \textbf{\bibinfo{volume}{74}},
  \bibinfo{pages}{124024} (\bibinfo{year}{2006}).

\bibitem[{\citenamefont{{Kastaun}}(2008)}]{2008arXiv0804.1151K}
\bibinfo{author}{\bibfnamefont{W.}~\bibnamefont{{Kastaun}}},
  \bibinfo{journal}{ArXiv e-prints} \textbf{\bibinfo{volume}{804}}
  (\bibinfo{year}{2008}).

\bibitem[{\citenamefont{{Bernuzzi} et~al.}(2008)\citenamefont{{Bernuzzi},
  {Nagar}, and {de Pietri}}}]{2008PhRvD..77d4042B}
\bibinfo{author}{\bibfnamefont{S.}~\bibnamefont{{Bernuzzi}}},
  \bibinfo{author}{\bibfnamefont{A.}~\bibnamefont{{Nagar}}}, \bibnamefont{and}
  \bibinfo{author}{\bibfnamefont{R.}~\bibnamefont{{de Pietri}}},
  \bibinfo{journal}{\prd} \textbf{\bibinfo{volume}{77}},
  \bibinfo{pages}{044042} (\bibinfo{year}{2008}).

\bibitem[{\citenamefont{{Bernuzzi} and {Nagar}}(2008)}]{2008arXiv0803.3804B}
\bibinfo{author}{\bibfnamefont{S.}~\bibnamefont{{Bernuzzi}}} \bibnamefont{and}
  \bibinfo{author}{\bibfnamefont{A.}~\bibnamefont{{Nagar}}},
  \bibinfo{journal}{ArXiv e-prints} \textbf{\bibinfo{volume}{803}}
  (\bibinfo{year}{2008}).

\bibitem[{\citenamefont{{Centrella} et~al.}(2001)\citenamefont{{Centrella},
  {New}, {Lowe}, and {Brown}}}]{2001ApJ...550L.193C}
\bibinfo{author}{\bibfnamefont{J.~M.} \bibnamefont{{Centrella}}},
  \bibinfo{author}{\bibfnamefont{K.~C.~B.} \bibnamefont{{New}}},
  \bibinfo{author}{\bibfnamefont{L.~L.} \bibnamefont{{Lowe}}},
  \bibnamefont{and} \bibinfo{author}{\bibfnamefont{J.~D.}
  \bibnamefont{{Brown}}}, \bibinfo{journal}{\apjl}
  \textbf{\bibinfo{volume}{550}}, \bibinfo{pages}{L193} (\bibinfo{year}{2001}).

\bibitem[{\citenamefont{{Shibata} et~al.}(2002)\citenamefont{{Shibata},
  {Karino}, and {Eriguchi}}}]{2002MNRAS.334L..27S}
\bibinfo{author}{\bibfnamefont{M.}~\bibnamefont{{Shibata}}},
  \bibinfo{author}{\bibfnamefont{S.}~\bibnamefont{{Karino}}}, \bibnamefont{and}
  \bibinfo{author}{\bibfnamefont{Y.}~\bibnamefont{{Eriguchi}}},
  \bibinfo{journal}{\mnras} \textbf{\bibinfo{volume}{334}},
  \bibinfo{pages}{L27} (\bibinfo{year}{2002}).

\bibitem[{\citenamefont{{Dimmelmeier} et~al.}(2002)\citenamefont{{Dimmelmeier},
  {Font}, and {M{\"u}ller}}}]{Dimmelmeier:2002bm}
\bibinfo{author}{\bibfnamefont{H.}~\bibnamefont{{Dimmelmeier}}},
  \bibinfo{author}{\bibfnamefont{J.~A.} \bibnamefont{{Font}}},
  \bibnamefont{and}
  \bibinfo{author}{\bibfnamefont{E.}~\bibnamefont{{M{\"u}ller}}},
  \bibinfo{journal}{\aap} \textbf{\bibinfo{volume}{393}}, \bibinfo{pages}{523}
  (\bibinfo{year}{2002}).

\bibitem[{\citenamefont{{Stavridis} et~al.}(2007)\citenamefont{{Stavridis},
  {Passamonti}, and {Kokkotas}}}]{2007PhRvD..75f4019S}
\bibinfo{author}{\bibfnamefont{A.}~\bibnamefont{{Stavridis}}},
  \bibinfo{author}{\bibfnamefont{A.}~\bibnamefont{{Passamonti}}},
  \bibnamefont{and}
  \bibinfo{author}{\bibfnamefont{K.}~\bibnamefont{{Kokkotas}}},
  \bibinfo{journal}{\prd} \textbf{\bibinfo{volume}{75}},
  \bibinfo{pages}{064019} (\bibinfo{year}{2007}).

\bibitem[{\citenamefont{{Passamonti}
  et~al.}(2008{\natexlab{a}})\citenamefont{{Passamonti}, {Stavridis}, and
  {Kokkotas}}}]{2008PhRvD..77b4029P}
\bibinfo{author}{\bibfnamefont{A.}~\bibnamefont{{Passamonti}}},
  \bibinfo{author}{\bibfnamefont{A.}~\bibnamefont{{Stavridis}}},
  \bibnamefont{and} \bibinfo{author}{\bibfnamefont{K.~D.}
  \bibnamefont{{Kokkotas}}}, \bibinfo{journal}{\prd}
  \textbf{\bibinfo{volume}{77}}, \bibinfo{pages}{024029}
  (\bibinfo{year}{2008}{\natexlab{a}}).

\bibitem[{\citenamefont{{Yoshida} et~al.}(2002)\citenamefont{{Yoshida},
  {Rezzolla}, {Karino}, and {Eriguchi}}}]{2002ApJ...568L..41Y}
\bibinfo{author}{\bibfnamefont{S.}~\bibnamefont{{Yoshida}}},
  \bibinfo{author}{\bibfnamefont{L.}~\bibnamefont{{Rezzolla}}},
  \bibinfo{author}{\bibfnamefont{S.}~\bibnamefont{{Karino}}}, \bibnamefont{and}
  \bibinfo{author}{\bibfnamefont{Y.}~\bibnamefont{{Eriguchi}}},
  \bibinfo{journal}{\apjl} \textbf{\bibinfo{volume}{568}}, \bibinfo{pages}{L41}
  (\bibinfo{year}{2002}).

\bibitem[{\citenamefont{{Boutloukos} and
  {Nollert}}(2007)}]{2007PhRvD..75d3007B}
\bibinfo{author}{\bibfnamefont{S.}~\bibnamefont{{Boutloukos}}}
  \bibnamefont{and} \bibinfo{author}{\bibfnamefont{H.-P.}
  \bibnamefont{{Nollert}}}, \bibinfo{journal}{\prd}
  \textbf{\bibinfo{volume}{75}}, \bibinfo{pages}{043007}
  (\bibinfo{year}{2007}).

\bibitem[{\citenamefont{{Dimmelmeier} et~al.}(2006)\citenamefont{{Dimmelmeier},
  {Stergioulas}, and {Font}}}]{2006MNRAS.368.1609D}
\bibinfo{author}{\bibfnamefont{H.}~\bibnamefont{{Dimmelmeier}}},
  \bibinfo{author}{\bibfnamefont{N.}~\bibnamefont{{Stergioulas}}},
  \bibnamefont{and} \bibinfo{author}{\bibfnamefont{J.~A.}
  \bibnamefont{{Font}}}, \bibinfo{journal}{\mnras}
  \textbf{\bibinfo{volume}{368}}, \bibinfo{pages}{1609} (\bibinfo{year}{2006}).

\bibitem[{\citenamefont{{Jones} et~al.}(2002)\citenamefont{{Jones},
  {Andersson}, and {Stergioulas}}}]{2002MNRAS.334..933J}
\bibinfo{author}{\bibfnamefont{D.~I.} \bibnamefont{{Jones}}},
  \bibinfo{author}{\bibfnamefont{N.}~\bibnamefont{{Andersson}}},
  \bibnamefont{and}
  \bibinfo{author}{\bibfnamefont{N.}~\bibnamefont{{Stergioulas}}},
  \bibinfo{journal}{\mnras} \textbf{\bibinfo{volume}{334}},
  \bibinfo{pages}{933} (\bibinfo{year}{2002}).

\bibitem[{\citenamefont{{Passamonti}
  et~al.}(2008{\natexlab{b}})\citenamefont{{Passamonti}, {Haskell},
  {Andersson}, {Jones}, and {Hawke}}}]{Passamonti:2008nx}
\bibinfo{author}{\bibfnamefont{A.}~\bibnamefont{{Passamonti}}},
  \bibinfo{author}{\bibfnamefont{B.}~\bibnamefont{{Haskell}}},
  \bibinfo{author}{\bibfnamefont{N.}~\bibnamefont{{Andersson}}},
  \bibinfo{author}{\bibfnamefont{D.~I.} \bibnamefont{{Jones}}},
  \bibnamefont{and} \bibinfo{author}{\bibfnamefont{I.}~\bibnamefont{{Hawke}}},
  \bibinfo{journal}{ArXiv e-prints} \textbf{\bibinfo{volume}{807}}
  (\bibinfo{year}{2008}{\natexlab{b}}).

\bibitem[{\citenamefont{{Ferrari} et~al.}(2007)\citenamefont{{Ferrari},
  {Gualtieri}, and {Marassi}}}]{2007PhRvD..76j4033F}
\bibinfo{author}{\bibfnamefont{V.}~\bibnamefont{{Ferrari}}},
  \bibinfo{author}{\bibfnamefont{L.}~\bibnamefont{{Gualtieri}}},
  \bibnamefont{and}
  \bibinfo{author}{\bibfnamefont{S.}~\bibnamefont{{Marassi}}},
  \bibinfo{journal}{\prd} \textbf{\bibinfo{volume}{76}},
  \bibinfo{pages}{104033} (\bibinfo{year}{2007}).

\bibitem[{\citenamefont{{Kojima}}(1998)}]{1998MNRAS.293...49K}
\bibinfo{author}{\bibfnamefont{Y.}~\bibnamefont{{Kojima}}},
  \bibinfo{journal}{\mnras} \textbf{\bibinfo{volume}{293}}, \bibinfo{pages}{49}
  (\bibinfo{year}{1998}).

\bibitem[{\citenamefont{Beyer and Kokkotas}(1999)}]{Beyer:1999te}
\bibinfo{author}{\bibfnamefont{H.~R.} \bibnamefont{Beyer}} \bibnamefont{and}
  \bibinfo{author}{\bibfnamefont{K.~D.} \bibnamefont{Kokkotas}},
  \bibinfo{journal}{\mnras} \textbf{\bibinfo{volume}{308}},
  \bibinfo{pages}{745} (\bibinfo{year}{1999}).

\bibitem[{\citenamefont{{Ruoff} and {Kokkotas}}(2002)}]{2002MNRAS.330.1027R}
\bibinfo{author}{\bibfnamefont{J.}~\bibnamefont{{Ruoff}}} \bibnamefont{and}
  \bibinfo{author}{\bibfnamefont{K.~D.} \bibnamefont{{Kokkotas}}},
  \bibinfo{journal}{\mnras} \textbf{\bibinfo{volume}{330}},
  \bibinfo{pages}{1027} (\bibinfo{year}{2002}).

\bibitem[{\citenamefont{{Ruoff} et~al.}(2003)\citenamefont{{Ruoff},
  {Stavridis}, and {Kokkotas}}}]{2003MNRAS.339.1170R}
\bibinfo{author}{\bibfnamefont{J.}~\bibnamefont{{Ruoff}}},
  \bibinfo{author}{\bibfnamefont{A.}~\bibnamefont{{Stavridis}}},
  \bibnamefont{and} \bibinfo{author}{\bibfnamefont{K.~D.}
  \bibnamefont{{Kokkotas}}}, \bibinfo{journal}{\mnras}
  \textbf{\bibinfo{volume}{339}}, \bibinfo{pages}{1170} (\bibinfo{year}{2003}).

\bibitem[{\citenamefont{{Ansorg} et~al.}(2003)\citenamefont{{Ansorg},
  {Kleinw{\"a}chter}, and {Meinel}}}]{Ansorg:2003mz}
\bibinfo{author}{\bibfnamefont{M.}~\bibnamefont{{Ansorg}}},
  \bibinfo{author}{\bibfnamefont{A.}~\bibnamefont{{Kleinw{\"a}chter}}},
  \bibnamefont{and} \bibinfo{author}{\bibfnamefont{R.}~\bibnamefont{{Meinel}}},
  \bibinfo{journal}{\aap} \textbf{\bibinfo{volume}{405}}, \bibinfo{pages}{711}
  (\bibinfo{year}{2003}).

\bibitem[{\citenamefont{Kojima}(1992)}]{Kojima:1992kj}
\bibinfo{author}{\bibfnamefont{Y.}~\bibnamefont{Kojima}},
  \bibinfo{journal}{Phys. Rev.} \textbf{\bibinfo{volume}{D46}},
  \bibinfo{pages}{4289} (\bibinfo{year}{1992}).

\bibitem[{\citenamefont{{Stavridis} and
  {Kokkotas}}(2005)}]{2005IJMPD..14..543S}
\bibinfo{author}{\bibfnamefont{A.}~\bibnamefont{{Stavridis}}} \bibnamefont{and}
  \bibinfo{author}{\bibfnamefont{K.~D.} \bibnamefont{{Kokkotas}}},
  \bibinfo{journal}{International Journal of Modern Physics D}
  \textbf{\bibinfo{volume}{14}}, \bibinfo{pages}{543} (\bibinfo{year}{2005}).

\bibitem[{\citenamefont{Gustafsson et~al.}(1995)\citenamefont{Gustafsson,
  Kreiss, and Oliger}}]{Gustafsson:1995yq}
\bibinfo{author}{\bibfnamefont{B.}~\bibnamefont{Gustafsson}},
  \bibinfo{author}{\bibfnamefont{H.-O.} \bibnamefont{Kreiss}},
  \bibnamefont{and} \bibinfo{author}{\bibfnamefont{J.}~\bibnamefont{Oliger}},
  \emph{\bibinfo{title}{Time dependent problems and difference methods}}
  (\bibinfo{publisher}{Wiley}, \bibinfo{year}{1995}).

\bibitem[{\citenamefont{{Chandrasekhar}}(1970)}]{1970PhRvL..24..611C}
\bibinfo{author}{\bibfnamefont{S.}~\bibnamefont{{Chandrasekhar}}},
  \bibinfo{journal}{Physical Review Letters} \textbf{\bibinfo{volume}{24}},
  \bibinfo{pages}{611} (\bibinfo{year}{1970}).

\bibitem[{\citenamefont{{Friedman} and {Schutz}}(1978)}]{1978ApJ...222..281F}
\bibinfo{author}{\bibfnamefont{J.~L.} \bibnamefont{{Friedman}}}
  \bibnamefont{and} \bibinfo{author}{\bibfnamefont{B.~F.}
  \bibnamefont{{Schutz}}}, \bibinfo{journal}{\apj}
  \textbf{\bibinfo{volume}{222}}, \bibinfo{pages}{281} (\bibinfo{year}{1978}).

\bibitem[{\citenamefont{{Lockitch} and {Friedman}}(1999)}]{1999ApJ...521..764L}
\bibinfo{author}{\bibfnamefont{K.~H.} \bibnamefont{{Lockitch}}}
  \bibnamefont{and} \bibinfo{author}{\bibfnamefont{J.~L.}
  \bibnamefont{{Friedman}}}, \bibinfo{journal}{\apj}
  \textbf{\bibinfo{volume}{521}}, \bibinfo{pages}{764} (\bibinfo{year}{1999}).

\bibitem[{\citenamefont{{Lockitch} et~al.}(2003)\citenamefont{{Lockitch},
  {Friedman}, and {Andersson}}}]{Lockitch:2003jt}
\bibinfo{author}{\bibfnamefont{K.~H.} \bibnamefont{{Lockitch}}},
  \bibinfo{author}{\bibfnamefont{J.~L.} \bibnamefont{{Friedman}}},
  \bibnamefont{and}
  \bibinfo{author}{\bibfnamefont{N.}~\bibnamefont{{Andersson}}},
  \bibinfo{journal}{\prd} \textbf{\bibinfo{volume}{68}},
  \bibinfo{pages}{124010} (\bibinfo{year}{2003}).

\bibitem[{\citenamefont{Boutloukos}(2006)}]{Boutloukos:2006dz}
\bibinfo{author}{\bibfnamefont{S.}~\bibnamefont{Boutloukos}}, Ph.D. thesis,
  \bibinfo{school}{Eberhard-Karls-Universit\"{a}t T\"{u}bingen}
  (\bibinfo{year}{2006}).

\bibitem[{\citenamefont{Stergioulas and Font}(2001)}]{Stergioulas:2000vs}
\bibinfo{author}{\bibfnamefont{N.}~\bibnamefont{Stergioulas}} \bibnamefont{and}
  \bibinfo{author}{\bibfnamefont{J.~A.} \bibnamefont{Font}},
  \bibinfo{journal}{Phys. Rev. Lett.} \textbf{\bibinfo{volume}{86}},
  \bibinfo{pages}{1148} (\bibinfo{year}{2001}).

\bibitem[{\citenamefont{Yoshida et~al.}(2005)\citenamefont{Yoshida, Yoshida,
  and Eriguchi}}]{Yoshida:2004gk}
\bibinfo{author}{\bibfnamefont{S.}~\bibnamefont{Yoshida}},
  \bibinfo{author}{\bibfnamefont{S.}~\bibnamefont{Yoshida}}, \bibnamefont{and}
  \bibinfo{author}{\bibfnamefont{Y.}~\bibnamefont{Eriguchi}},
  \bibinfo{journal}{Mon. Not. Roy. Astron. Soc.}
  \textbf{\bibinfo{volume}{356}}, \bibinfo{pages}{217} (\bibinfo{year}{2005}).

\bibitem[{\citenamefont{Stergioulas and Friedman}(1998)}]{Stergioulas:1997ja}
\bibinfo{author}{\bibfnamefont{N.}~\bibnamefont{Stergioulas}} \bibnamefont{and}
  \bibinfo{author}{\bibfnamefont{J.~L.} \bibnamefont{Friedman}},
  \bibinfo{journal}{Astrophys. J.} \textbf{\bibinfo{volume}{492}},
  \bibinfo{pages}{301} (\bibinfo{year}{1998}).

\bibitem[{\citenamefont{Karino et~al.}(2001)\citenamefont{Karino, Yoshida, and
  Eriguchi}}]{PhysRevD.64.024003}
\bibinfo{author}{\bibfnamefont{S.}~\bibnamefont{Karino}},
  \bibinfo{author}{\bibfnamefont{S.}~\bibnamefont{Yoshida}}, \bibnamefont{and}
  \bibinfo{author}{\bibfnamefont{Y.}~\bibnamefont{Eriguchi}},
  \bibinfo{journal}{Phys. Rev. D} \textbf{\bibinfo{volume}{64}},
  \bibinfo{pages}{024003} (\bibinfo{year}{2001}).

\end{thebibliography}
%%%%%%%%%%%%%%%%%%%%%%%%%%%%%%%%%%%%%%%%%%%%%%%%%%%%%%%%%%%%%%%%%%%%%%%%%%%%%%%%
\end{document}